\documentclass[reprint,NumberedRefs]{JASA}

\usepackage{siunitx}
\usepackage{times}
\usepackage{mathptmx}

\begin{document}

\title[JASA/Non-Hermitian acoustic metamaterial]{Realizing non-Hermitian tunneling phenomena using non-reciprocal active acoustic metamaterials}

\author{Felix Langfeldt}
\email{F.Langfeldt@soton.ac.uk}
\author{Joe Tan}
\affiliation{Institute of Sound and Vibration Research, University of Southampton, University Road, Highfield, Southampton, SO17 1BJ, United Kingdom}

\author{Sayan Jana}

\author{Lea Sirota}
\email{leabeilkin@tauex.tau.ac.il}
\affiliation{School of Mechanical Engineering, Tel Aviv University, Tel Aviv 69978, Israel}

\begin{abstract}

Non-reciprocal systems have been shown to exhibit various interesting wave phenomena, such as the non-Hermitian skin effect, which causes accumulation of modes at boundaries. Recent research on discrete systems showed that this effect can pose a barrier for waves hitting an interface between reciprocal and non-reciprocal systems. Under certain conditions, however, waves can tunnel through this barrier, similar to the tunneling of particles in quantum mechanics. This work proposes and investigates an active acoustic metamaterial design to realize this tunneling phenomenon in the acoustical wave domain. The metamaterial consists of an acoustic waveguide with microphones and loudspeakers embedded in its wall. Starting from a purely discrete non-Hermitian lattice model of the system, a hybrid continuous-discrete acoustic model is derived, resulting in distributed feedback control laws to realize the desired behavior for acoustic waves. The proposed control laws are validated using frequency and time domain finite element method simulations, which include lumped electro-acoustic loudspeaker models. Additionally, an experimental demonstration is performed using a waveguide with embedded active unit cells and a digital implementation of the control laws. In both the simulations and experiments the tunneling phenomenon is successfully observed.

\end{abstract}

\maketitle

\section{Introduction}

Acoustic metamaterials are artificial structures, often involving a periodic assembly of unit cells, which are engineered to control sound waves in unconventional ways. Unlike natural materials, the metamaterials exhibit properties that arise from their architectured couplings rather than the material composition \cite{liu2000locally, cummer2016controlling}, making them ideal for applications that require precise control over acoustic wave propagation.
Notable properties include, for example, effective negative constitutive parameters and refractive indices in the subwavelength regime, enabling applications such as acoustic cloaking, focusing, or lensing\cite{cummer2007one,craster2012acoustic,norris2015acoustic}.

Other properties are based on non-Hermitian physics, originally associated with quantum mechanics\cite{ashida2020non}. In non-Hermitian systems, the spectrum is typically complex-valued, which provides new insights into the originally Hermitian concepts of topological invariants, bulk-boundary correspondence and its failure, as well as the topological protection of boundary modes\cite{yao2018edge}.
For example, by balancing between gain and loss, parity-time symmetry\cite{bender1998real} can be obtained. The properties of the associated exceptional points were utilized for unidirectional acoustic invisibility, cloaking, coherent absorption, and more \cite{zhu2014pt,fleury2016parity,shi2016accessing,achilleos2017non,gu2021acoustic,gu2021controlling,xu2022exceptional}.

An aspect of non-Hermitian physics that has gained a particularly enhanced interest in recent years, and has been employed in acoustics, is nonreciprocity. 
In certain nonreciprocal acoustic systems, wave propagation is enhanced in one direction while being weakened in the opposite. This unidirectional behavior is linked to the well-known non-Hermitian skin effect, wherein wave energy becomes localized at the boundaries of a system. The non-Hermitian skin effect is currently being explored for its potential to design highly directional acoustic devices and waveguides
\cite{lee2019anatomy,goldsberry2019non,nassar2020nonreciprocity,scheibner2020non,rosa2020dynamics,helbig2020generalized,rasmussen2021acoustic,zhang2021acoustic,song2022skin,zhang2022non,wen2024nonreciprocal}. 

Some of the metamaterial properties, such as effective negative parameters (in a finite frequency range), or loss, can be obtained passively using engineering of the unit cell geometry, adding resonant inclusions, dissipative inclusions, and so on. Other properties, however, such as all-frequency gain or nonreciprocity, usually require active components. 
Active metamaterials in diverse fields enhance the capabilities of their passive counterparts by allowing for real-time manipulation of the underlying wave propagation. This dynamic control can be achieved using distributed feedback loops, where actuators inject energy into the system, based on sensor measurements processed by micro-controllers, or by inherent active feedback elements as in electric circuits\cite{helbig2020generalized,halder2024circuit,benisty2025controlled}. 

This work focuses on distributed feedback-based acoustic metamaterials. Therein, loudspeakers and microphones are, respectively, actuators and sensors that can tune the existing properties to adapt to changing environmental conditions, as well as to create new otherwise inaccessible structural couplings, such as nonreciprocity\cite{zhang2021acoustic,geib2021tunable,tan2022realisation,langfeldt2023controlling,wen2023acoustic,guo2023observation,maddi2024exact,tan2024realisation,kovacevich2024stability,wang2025supersonic}.
In particular, a recently reported intriguing wave dynamics phenomenon is considered here, which combines the non-Hermitian skin effect with tunneling-like behavior to create a dark/quiet zone in the system's interior\cite{jana2025invisible}.

\begin{figure}
\figline{
        \fig{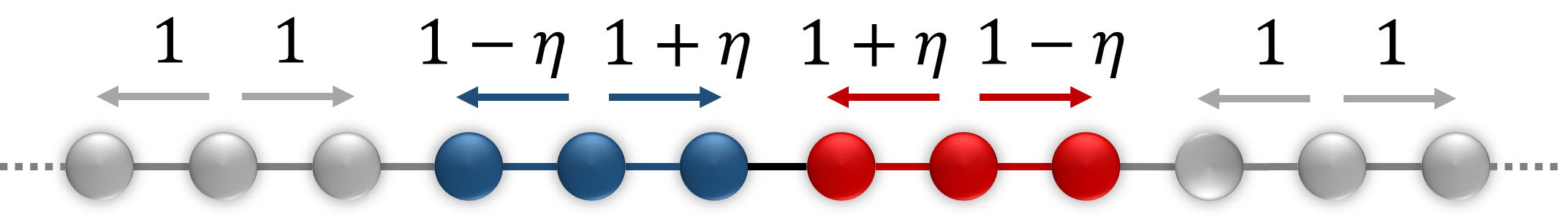}{8.3cm}{(a)}
        \label{fig:quantum_chain}
}
\figline{
        \fig{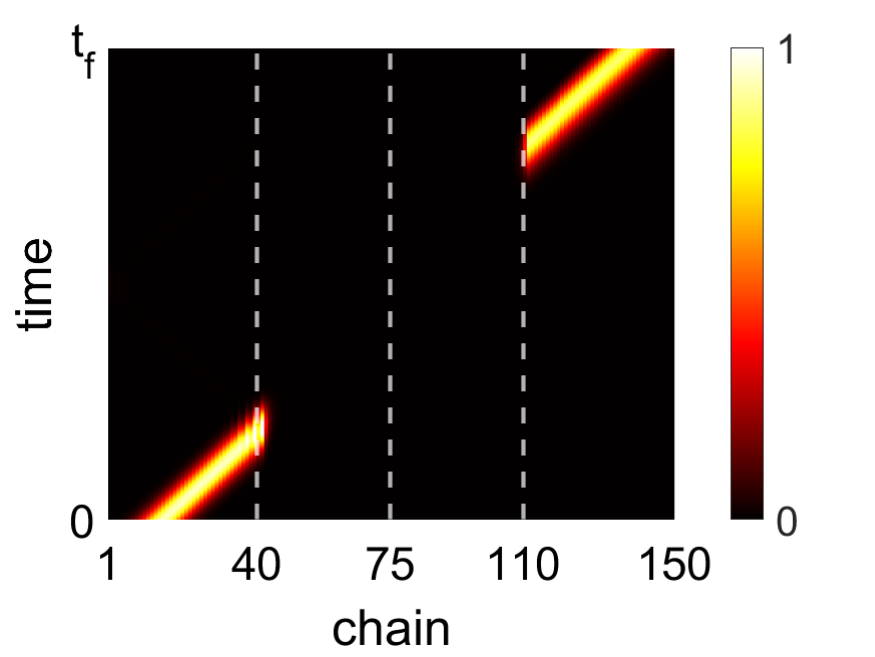}{4.4cm}{(b)}
        \label{fig:quantum_tunnel_q}
        \fig{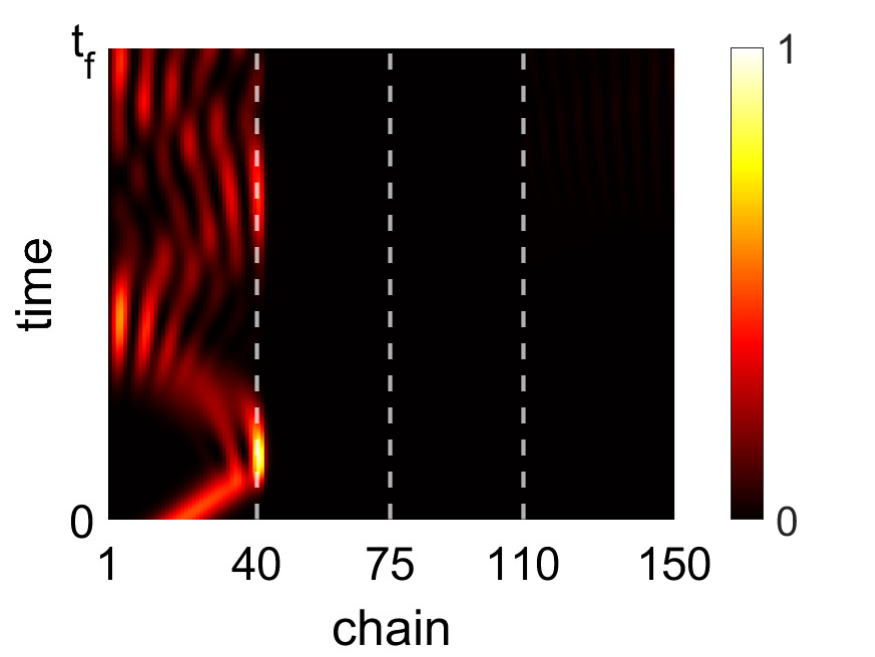}{4.4cm}{(c)}
        \label{fig:quantum_tunnel_refl}
}
\caption{The quantum analogy of the tunneling phenomenon. (a) Lattice schematic partitioned into two Hermitian sections (gray) and two non-Hermitian sections (blue, red). 
(b)-(c) Gaussian wavepacket evolution in a 150 sites lattice with 70 non-Hermitian sites, and $\eta=0.5$. (b) Wavepacket corresponding to an energy within the propagation window $|E|<2\sqrt{1-\eta^2}$, demonstrating tunneling-like transmission through the interface. (c) Wavepacket corresponding to an energy outside the window, demonstrating total reflection.}
\label{fig:quantum}
\end{figure}

The phenomenon, dubbed non-Hermitian tunneling, was featured in a purely discrete system---a lattice, in the quantum realm. 
Specifically, it was shown that while the non-Hermitian skin effect is obtained for nonreciprocity along an entire non-Hermitian lattice, or a chain in one dimension, a tunneling-like phenomenon emerges when placing two nonreciprocal chains, as depicted by blue and red in \autoref{fig:quantum_chain}, as an interface between reciprocal chains, depicted by gray. In the quantum system, the nonreciprocity of the blue chain arises since the electron creation operator $\alpha^\dagger$ at each site is coupled to the annihilation operator $\alpha$ of its nearest neighbor with a stronger coupling of $1+\eta$ to the right, and with a weaker coupling of $1-\eta$ to the left, where $\eta\in (0,1)$. For the red section this definition is flipped.
In the gray sections, the coupling is equal in both directions. Due to the underlying structural nonreciprocity, the blue and red sections are governed by non-Hermitian Hamiltonians of the Hatano-Nelson type\citep{hatano1996localization} with $H=\textstyle{\sum}_{j}\left(1+\eta\right)\alpha_{j}^{\dagger}\alpha_{j+1}+\left(1-\eta\right)\alpha_{j+1}^{\dagger}\alpha_{j}$ for the blue section, and with $1+\eta$ and $1-\eta$ interchanged for the red section.

It was discovered that for the energy window $|E|<2\sqrt{1-\eta^2}$, a wavepacket $\Psi$ that propagates via Schr\"odinger dynamics $\mathrm{i}\hbar \mathrm d \Psi(t) / \mathrm d t=H\Psi(t)$ along a Hermitian section and hits the non-Hermitian interface seemingly disappears, and reemerges on the other side of the interface at a later time, as demonstrated by the simulation in \autoref{fig:quantum_tunnel_q}. 
This unique phenomenon portrays a similar effect as if the wave invisibly tunneled through the interface. 
Outside this energy window the wave is fully reflected by the interface, see \autoref{fig:quantum_tunnel_refl}. 

The aim of the research in this contribution is to realize an acoustic analogy of this phenomenon in a one-dimensional waveguide, where the dark interface of \autoref{fig:quantum_tunnel_q} will be mapped to an artificial interface in the waveguide, through which the sound wave tunnels, creating a quiet region.
The nonreciprocal interface of \autoref{fig:quantum_chain} will be implemented using active feedback elements embedded in the waveguide. The key challenges addressed by this work are: (i) to correctly map the purely discrete model onto the continuous waveguide system, (ii) to operate the distributed feedback mechanism without altering the waveguide geometry or blocking the air passage (so to enable versatility of applications), and (iii) to design the control laws that balance between the quiet region length and the tunneling strength for stable tunneling dynamics.

The paper is organized as follows: In \autoref{sec:metamaterial_design}, the design of the proposed active acoustic metamaterial is derived theoretically, via a mapping of the purely discrete model to the acoustic domain. Using lumped element models for the control sources, the control laws for creating the desired tunneling behavior within the waveguide are derived.
\autoref{sec:numerical} then presents a numerical analysis of the proposed system, comparing and validating the tunneling of acoustic waves using theoretical models and finite element method simulations.
\autoref{sec:experiment} describes the experimental realization of the active acoustic metamaterial, and shows measurements results that demonstrate the tunneling phenomenon for acoustic waves under realistic experimental conditions.
Finally, the key findings of this contribution are summarized and concluded in \autoref{sec:conclusion}.

\section{Active metamaterial design}
\label{sec:metamaterial_design}

\begin{figure*}
\figline{
        \fig{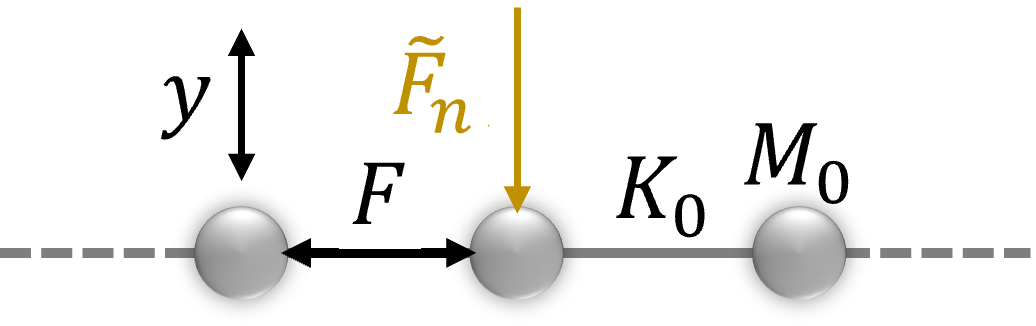}{4.7cm}{(a)}
        \label{fig:acoustic_waveguide_ctrl_short}
        \fig{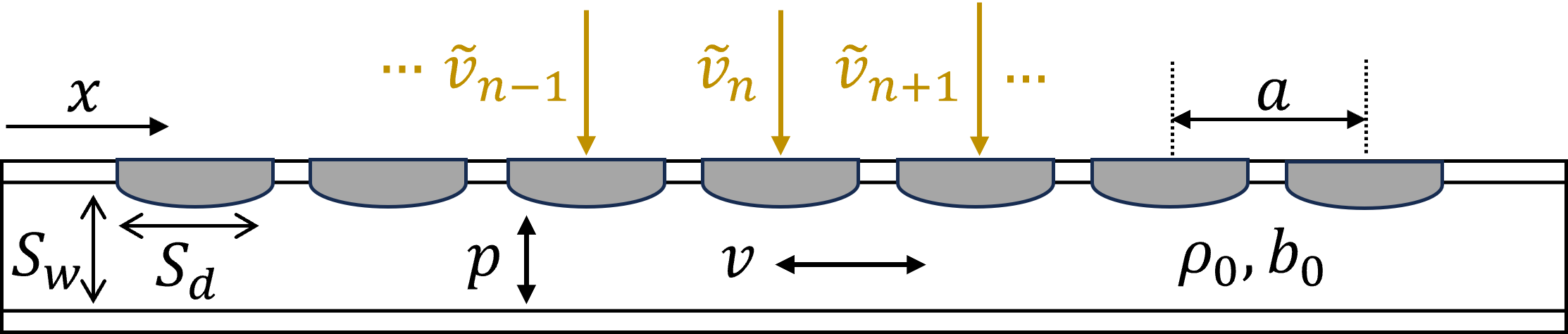}{9.9cm}{(c)}
        \label{fig:acoustic_waveguide_ctrl_a}
}
\figline{
        \fig{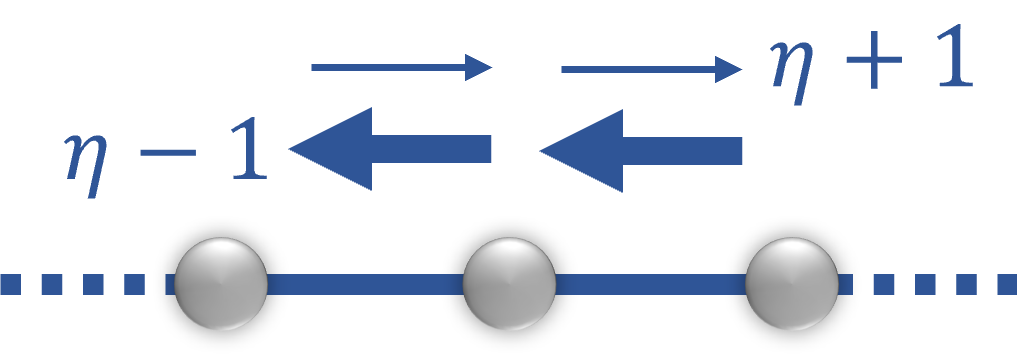}{4.7cm}{(b)}
        \label{fig:acoustic_discrete_cell_nonrecip}
        \fig{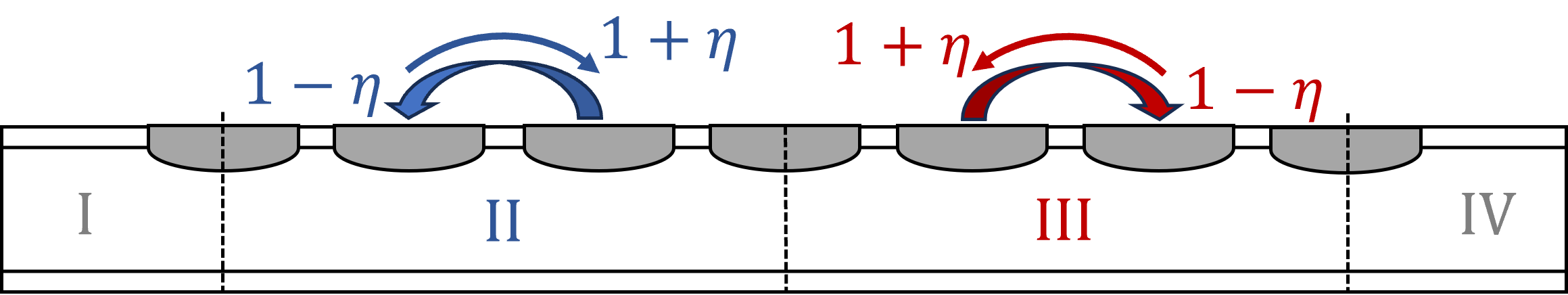}{9.9cm}{(d)}
        \label{fig:waveguide_CL_1}
}
    \caption{Model of the acoustic metamaterial. (a) The classical analogy lattice model, uncontrolled. (b) The closed loop nonreciprocal classical lattice. (c) The waveguide schematic, featuring an array of electroacoustic actuators (gray discs). (d) The waveguide schematic in closed loop, with the nonreciprocal couplings created by the controllers.}
    \label{fig:acoustic_waveguide}
\end{figure*}

To derive an acoustic analogy to the quantum system in \autoref{fig:quantum_chain}, a classical mass-spring lattice model (a chain in one dimension) is first considered. Therein, the electron hopping is mimicked by the vibration of the masses $M_0$, and the hopping strength is equivalent to the stiffness $K_0$ of the springs connecting the masses, as illustrated in \autoref{fig:acoustic_waveguide_ctrl_short}. 
To break the reciprocity in the chain, the vibration velocity $\mathrm d y/\mathrm d t$ of each mass is controlled in a distributed feedback loop by an external force $\widetilde{F}$. For the $n$-th mass, the control force depends on the velocities of the adjacent masses as
\begin{equation}   \label{eq:F_cont}
    \widetilde{F}_n=K_0\eta\int_0^t\left(\dfrac{\mathrm d y_{n+1}}{\mathrm d t}-\dfrac{\mathrm d y_{n-1}}{\mathrm d t}\right)\mathrm{d}t.
\end{equation}
Here, the nonreciprocity strength $\eta$ is taken positive for the blue chain and negative for the red. 
The resulting closed loop nonreciprocal blue chain is illustrated in \autoref{fig:acoustic_discrete_cell_nonrecip}. In this chain, the effective spring stiffness equals $K_0(1+\eta)$ to the right and $K_0(1-\eta)$ to the left, leading to directional wave dynamics with a preferred propagation to the left.

At the next stage, the classical nonreciprocal lattice model is mapped onto a continuous acoustic domain. 
An acoustic waveguide with cross-sectional area $S_w$ is considered, which enables the propagation of sound pressure waves $p$ through a fluid with mass density $\rho_0$ and bulk modulus $b_0$, as shown in \autoref{fig:acoustic_waveguide_ctrl_a}. Inward-facing active elements of area $S_d$ are attached to the waveguide wall with a periodic spacing $a$. The principal propagation axis is denoted $x$.
The pressure $p$ and the associated acoustic velocity $v$ in the waveguide can be effectively mapped to the velocities $\mathrm d y/\mathrm d t$ of the masses and internal spring forces $F$, respectively, via the constitutive equations
\begin{equation}   \label{eq:constitutive}
    \begin{cases} 
    \dfrac{\mathrm d y_{n+1}}{\mathrm d t}-\dfrac{\mathrm d y_{n}}{\mathrm d t}=-\dfrac{1}{K_0}\dfrac{\mathrm d F_{n+1}}{\mathrm d t} \\ 
    \dfrac{\mathrm d^2 y_{n}}{\mathrm d t^2}=-\dfrac{1}{M_0}\left(F_{n+1}-F_n\right)
    \end{cases} \quad  \leftrightarrow \quad \begin{cases} 
    \dfrac{\partial p}{\partial x}=-\rho_0 \dfrac{\partial v}{\partial t} \\ 
    \dfrac{\partial p}{\partial t}=-b_0 \dfrac{\partial v}{\partial x}
    \end{cases}.
\end{equation}
The masses $M_0$ and spring constants $K_0$ are then analogous to the bulk modulus and mass density of the fluid via $M_0\leftrightarrow aS_w/b_0$ and $K_0 \leftrightarrow S_w/(\rho_0a)$. 
In order to realize the required non-Hermitian couplings in the waveguide, the active elements in the waveguide walls are used to break the acoustic wave reciprocity. Each element at location $x_n$ generates an acoustic velocity input $\widetilde{v}_n$, so that the pressure field inside the waveguide is governed by
\begin{equation}  \label{eq:OL_field}
    \dfrac{\partial^2 p}{\partial t^2}=c^2 \dfrac{\partial^2 p}{\partial x^2}+b_0\beta\sum_n \dfrac{ \mathrm d \widetilde{v}_n}{\mathrm d t}\delta(x-x_n),
\end{equation}
with $c=\sqrt{b_0/\rho_0}$ being the speed of sound, and $\beta=S_d/S_w$. With the equivalence relationships in \autoref{eq:constitutive} and following the lattice control signals in \autoref{eq:F_cont}, the velocity control signals are given by
\begin{equation} \label{eq:v_n}
    \widetilde{v}_n=\frac{\eta}{\rho_0a\beta}\int_0^t \Delta p_n\;\mathrm{d}t,
\end{equation}
where
\begin{equation}  \label{eq:Delta_pn}
    \Delta p_n=\begin{cases} n_\mathrm{I/II}: \quad & p(x_{n+1},t)-p(x_n,t) \\ n\in\mathrm{II}: \quad & p(x_{n+1},t)-p(x_{n-1},t) \\ n_\mathrm{II/III}: \quad & p(x_{n+1},t)-2p(x_n,t)+p(x_{n-1},t) \\ n\in\mathrm{III}: \quad & p(x_{n-1},t)-p(x_{n+1},t) \\ n_\mathrm{III/VI}: \quad & p(x_{n-1},t)-p(x_n,t)
    \end{cases}.
\end{equation}
Here, I-IV represent the metamaterial sectioning according to \autoref{fig:waveguide_CL_1}, in which I and IV represent the Hermitian sections, whereas II and III represent the mirrored non-Hermitian interface sections---the analogy of the blue and red quantum chains in \autoref{fig:quantum_chain}. The transition cells between the sections are labeled by I/II, II/III, and III/IV.
The Hermitian sections are, therefore, given by a plain uncontrolled waveguide.
The control sources therefore induce in Sections II and III the required $1\pm\eta$ couplings between the sites
based on the real-time measurements of the pressure field responses in the current and neighboring cells, which can be done using microphones distributed along the waveguide.

\begin{figure}
\figline{
        \fig{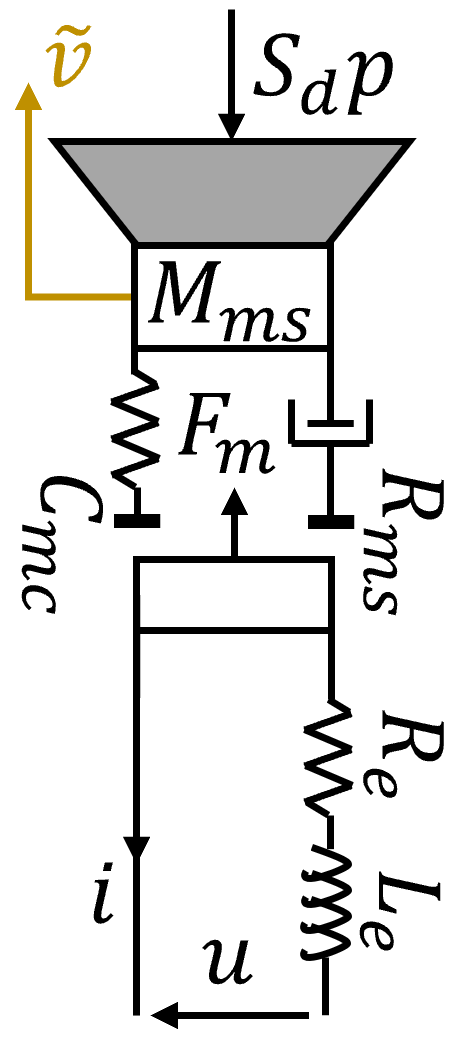}{2cm}{(a)}
        \label{fig:loudspeaker_sketch}
        \fig{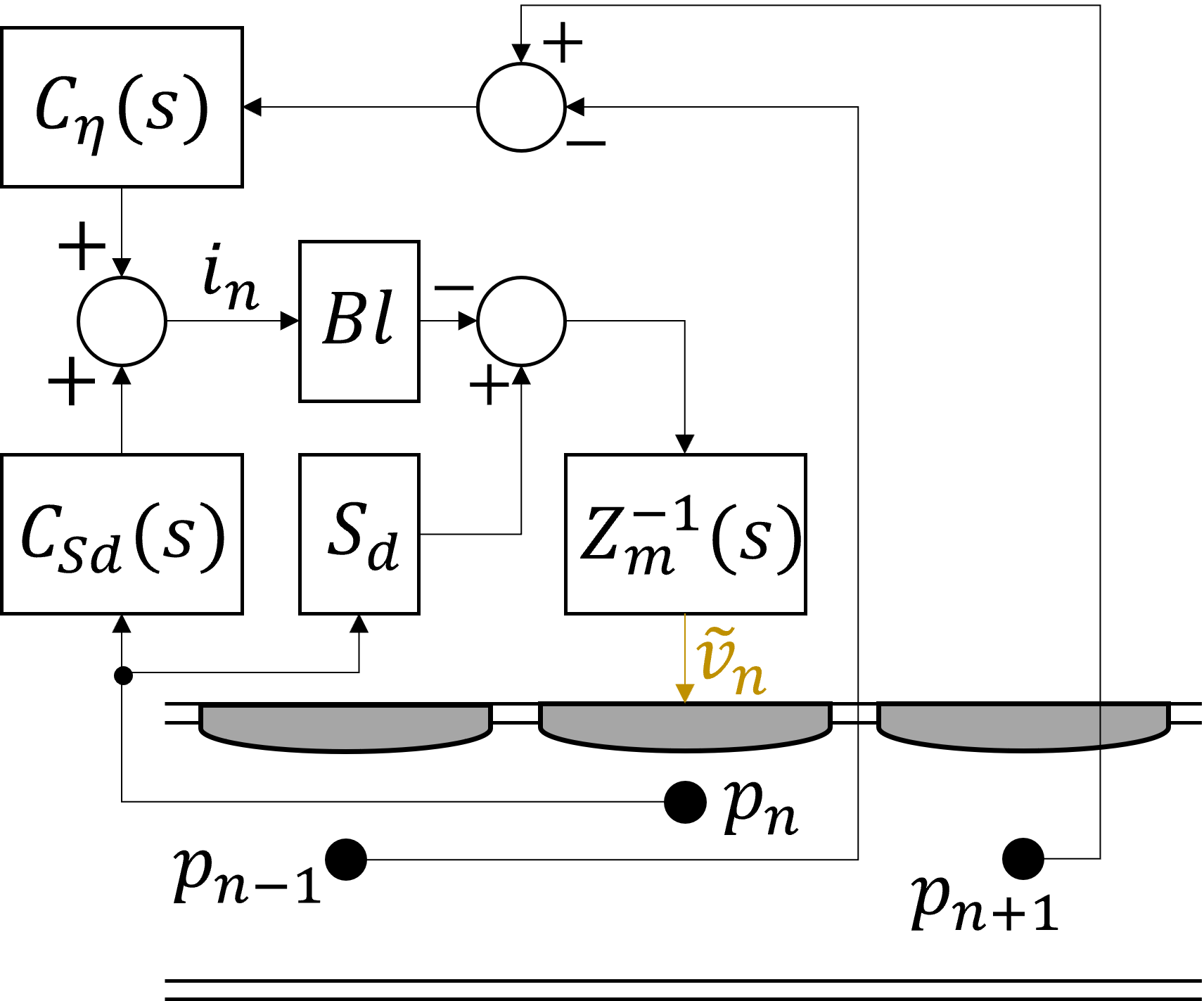}{5.5cm}{(b)}
        \label{fig:loudspeaker_ctrl}
}
    \caption{(a) Electroacoustic actuator model. (b) The controller block diagram for $n\in II$.}
    \label{fig:loudspeaker}
\end{figure}

In this work, the control sources are represented by electrodynamic loudspeakers with a closed back-cavity, as illustrated in \autoref{fig:loudspeaker_sketch}. At low frequencies, each loudspeaker can be approximated as a mass-spring-damper system, and the small displacement diaphragm response to an electric input signal can be described in the Laplace domain (with the Laplace variable $s$) via \cite{rivet2016broadband,geib2021tunable}
\begin{subequations} \label{eq:speaker_eqs}
\begin{align}
    Z_{mo}(s)\widetilde{v}_n(s)&=-S_dp_n(s)+Bli_n(s), \label{eq:speaker_eqs_i}  \\
 u_n(s)&=Z_{eb}(s)i_n(s)+Bl \widetilde{v}_n(s).  \label{eq:speaker_eqs_u}
\end{align} 
\end{subequations}
Here, $Z_{mo}(s)=M_{ms}s+R_{ms}+\frac{1}{C_{mc}s}$ and $Z_{eb}(s)=L_es+R_e$
are, respectively, the open circuit mechanical and the blocked electrical impedance of the loudspeaker. $M_{ms}$, $R_{ms}$, and $C_{mc}$ represent its moving mass, mechanical damping, and the total mechanical compliance. $S_d$ is the effective area of the diaphragm, with $p_n$ being the total sound pressure acting on it, which includes both the incident and scattered pressure. $\widetilde{v}_n$ is the vibration velocity of the speaker diaphragm, $i_n$ is the current in the voice coil, and $u_n$ is the input voltage between the electrical terminals. $Bl$ is the force factor of the speaker, $R_e$ is the DC resistance, and $L_e$ is the self-inductance of the voice coil. 
To avoid the impact of the coil inductance $L_e$ on the system stability, the loudspeakers were driven by current sources.
The corresponding current control commands are then given by
\begin{equation} \label{eq:u_rep_n}
    i_n(s)=C_{Sd} p_n(s)-C_\eta(s)\Delta p_n(s),
\end{equation}
where $\Delta p_n(s)$ is defined in \autoref{eq:Delta_pn}, and $C_{Sd}$ and $C_\eta(s)$ are the controllers for the loudspeaker self-dynamics cancellation and non-reciprocity realization, respectively. The controller block diagram is illustrated in \autoref{fig:loudspeaker_ctrl}. 
By substituting \autoref{eq:u_rep_n} into \autoref{eq:speaker_eqs_i}, and equating with \autoref{eq:v_n} and \autoref{eq:Delta_pn}, the controllers take the form
\begin{equation}   \label{eq:controllers}
        C_{Sd}(s)=\frac{S_d}{Bl}, \quad 
        C_\eta(s)=\frac{\eta Z_{mo}(s)}{\rho_0aBl\beta s}
        \text{.}
\end{equation}

\begin{figure*}
    \figline{
        \fig{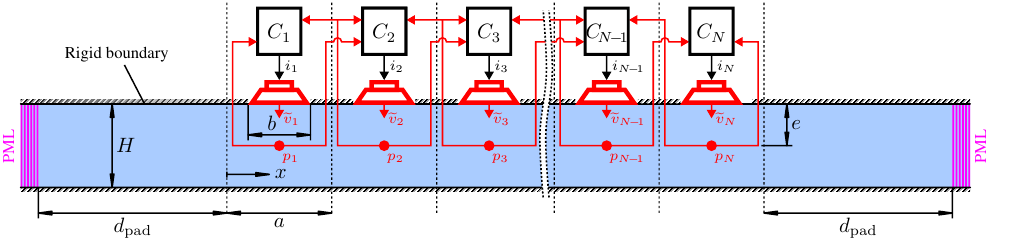}{6.2in}{}
        \label{fig:fem_overview}
    }
    \caption{Illustration of the two-dimensional finite element model setup of the waveguide with the proposed active acoustic metamaterial.}
    \label{fig:fem_setup}
\end{figure*}

\section{Numerical analysis}

\label{sec:numerical}

In this section the model for tunneling of acoustic waves described in \autoref{sec:metamaterial_design} is validated numerically. Simulations of the controlled lumped-parameter lattice model in \autoref{eq:F_cont} and \autoref{eq:constitutive} are compared to finite element (FE) simulations of the acoustic waveguide system controlled using realistic electroacoustic actuators according to the current-based control laws given in \autoref{eq:u_rep_n} and \autoref{eq:controllers}.

\subsection{Finite element model setup}
\label{sec:numerical_fem_setup}

The FE model of the proposed active metamaterial design has been developed to investigate the performance and stability of the metamaterial when including physical effects that cannot be captured by the one-dimensional lattice model (e.g., finite actuator size, microphone location, and higher-order acoustic waveguide modes). The FE model is two-dimensional and represents a cross section of the waveguide that has been used for the experimental demonstration (see \autoref{sec:experiment}). A representative sketch of the FE model, highlighting all relevant dimensions and boundary conditions, is shown in \autoref{fig:fem_setup}. \autoref{tbl:fem_parameters} provides an overview of the numerical values used for all geometrical and material parameters in the model.

\begin{table}
    \centering
    \caption{Parameters of the FE model of the waveguide with the proposed active metamaterial.}
    \label{tbl:fem_parameters}
    \begin{ruledtabular}
        \begin{tabular}{ccccccc}
            $H$ & $a$ & $b$ & $e$ & $d_\mathrm{pad}$ & $\rho_0$ & $c_0$ \\
            \hline

            40 & 50 & 19.5 & 20 & 10 & 1.2 & 343 \\
            
            \hline

            \si{\milli\meter} & \si{\milli\meter} & \si{\milli\meter} & \si{\milli\meter} & \si{\centi\meter} & \si{\kilo\gram\per\meter\cubed} & \si{\meter\per\second}  \\
        \end{tabular}
    \end{ruledtabular}
\end{table}

The waveguide with height $H$ was truncated at both ends by perfectly matched layers (PML) to fully absorb sound waves leaving the waveguide and minimize the impact of reflections on the simulation results. A padding distance of $d_\mathrm{pad}$ was used between the active metamaterial and both PML to reduce possible interactions between evanescent waves and the PML. The upper and lower walls of the waveguide were modeled as rigid walls, except for the control sources which were modeled as boundaries with a prescribed normal velocity $\widetilde{v}_n$.
To take into account the finite size of the control sources, the control source length $b$ was defined as $b=\beta H$, with $\beta=0.49$ as in the experiment (see \autoref{sec:experiment_setup}).
Point probes were used to extract the total acoustic pressure $p_n$ at the microphone locations. As in the experimental setup, each point probe was located a distance of $e=H/2$ in normal direction from the center of the corresponding control source.
The system was acoustically excited using a background plane acoustic wave signal propagating in positive $x$ direction. Depending on the study type, the background wave signal can, for example, be a sinusoidal wave (e.g., for frequency domain simulations) or a sine-modulated Gaussian pulse (for time domain simulations).

The FE model was spatially discretized using triangular quadratic Lagrange elements with a maximum element size of $\SI{4.2}{\milli\meter}$, ensuring at least six elements per wavelength for frequencies up to \SI{13.6}{\kilo\hertz}, which is well above the frequency range of interest for this study. To ensure that the simulation results are mesh-independent, a convergence study was performed by halving the element size ($\SI{2.1}{\milli\meter}$) for which the simulation results did not change significantly compared to the coarser grid.

For a more realistic representation of the control source dynamics, the normal velocity of each actuator $\widetilde{v}_n$ (as a response to a control current input $i_n$) was calculated using the lumped element model for electrodynamic loudspeaker drivers given in \autoref{eq:speaker_eqs} and the control laws from \autoref{eq:u_rep_n}.
The discretized FE model, the ODEs for the actuator dynamics, and the control laws were solved simultaneously, taking into account the full coupling between the acoustic pressure field $p$, the control source velocities $\widetilde{v}_n$, and the control currents $i_n$.
When solving the model in the frequency domain, all time derivatives were replaced by $\mathrm{i}\omega$, and the complex-valued results were obtained at each frequency using a direct linear solver.
For time domain simulations, the implicit generalized-$\alpha$ method \citep{chung1993genalph} was used for time-stepping with a constant time-step size of $\Delta t = \SI{4}{\micro\second}$. The suitability of the chosen time-step size was verified by performing additional simulations with a smaller time-step size (\SI{2}{\micro\second}), showing no significant difference compared to the results obtained with the chosen value for $\Delta t$.

\begin{table}
    \centering
    \caption{Thiele/Small paramters of the control sources used in the FE simulations in \autoref{sec:numerical_fem_setup} and the experiment in \autoref{sec:experiment_setup} (Peerless TC5FB00-04).}
    \label{tbl:ts_parameters}
    \begin{ruledtabular}
        \begin{tabular}{ccccccccc}
            $R_e$ & $L_e$ & $M_{ms}$ & $C_{ms}$ & $R_{ms}$ & $f_s$ & $Bl$ & $S_d$ \\
            \hline

            3.58 & 0.04 & 0.5 & 0.92 & 0.16 & 235 & 0.96 & 7.8 \\
            
            \hline

            \si{\ohm} & \si{\milli\henry} & \si{\gram} & \si{\milli\meter\per\newton} & \si{\kilo\gram\per\second} & \si{\hertz} & \si{\tesla\meter} & \si{\centi\meter\squared}  \\
        \end{tabular}
    \end{ruledtabular}
\end{table}

\subsection{Reflection coefficient and decay rate}
\label{sec:numerical_refl_decay}

\begin{figure}
\figcolumn{
        \fig{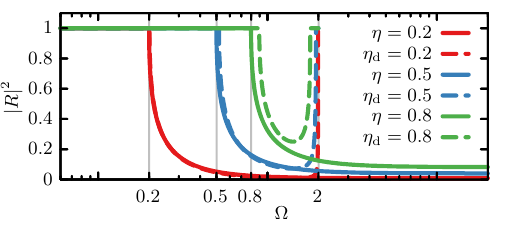}{2.9in}{(a)}
        \label{fig:R_theory}
        \fig{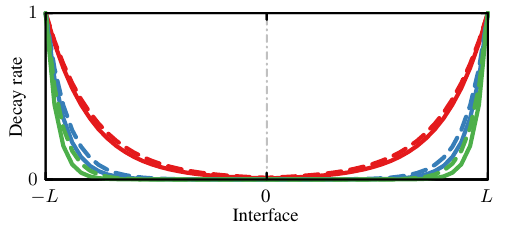}{2.9in}{(b)}
        \label{fig:decay_theory}
        \fig{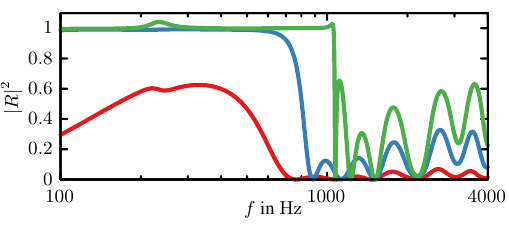}{2.9in}{(c)}
        \label{fig:R_fem_9}
        \fig{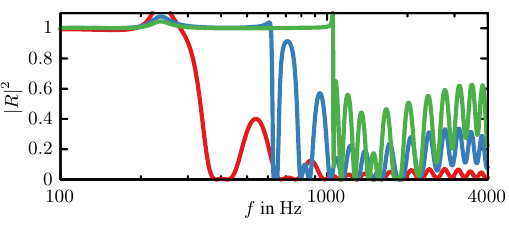}{2.9in}{(d)}
        \label{fig:R_fem_21}
    }
    \caption{(a) Reflection coefficient $R$ and (b) decay rate $q$ of the theoretical continuous (solid) and discrete (dashed) systems, obtained from \autoref{eq:R_theory}, for different values of $\eta$. (c)-(d) FE simulated reflection coefficient frequency responses for the active acoustic metamaterial in \autoref{fig:fem_setup} with (c) $N=9$ and (d) $N=21$ unit cells and different values of $\eta$.
    }
    \label{fig:fem_reflection}
\end{figure}

To demonstrate the tunneling phenomenon in the acoustic waveguide system, the conditions at which the phenomenon can take place under classical dynamics are derived first. 
Specifically, the reflection coefficient $R$ of waves incident towards the non-Hermitian interface from the Hermitian sections will be derived. 
In addition, the decay rate $q_d$ for the discrete and $q_c$ for the continuous system, which is the measure of the interface darkness, and indicates the tunneling invisibility level, is also derived.
Both a discrete system, such as a mass-spring lattice that is equivalent to the atomic lattice in \autoref{fig:quantum_chain}, and a continuous system, representing an effective material, which is obtained when the differences in \autoref{eq:Delta_pn} are treated as a first order spatial derivative added to the wave equation in \autoref{eq:OL_field}, resulting in $\partial^2 p /\partial t^2=c^2 \partial^2 p /\partial x^2 +2\eta\omega_0c \partial p /\partial x$, where $\omega_0=\sqrt{K_0/M_0}$, are considered. The waveguide model from \autoref{fig:fem_overview}, which is a hybrid continuous-discrete system, will be tested in conjunction with these two marginal cases.
The expressions of $R$, $q_d$, and $q_c$ take the form~\cite{jana2025invisible}
\begin{equation}
\begin{split}
   & R=\frac{\mathrm{i}\widehat{\Omega}-f_\eta(\widehat{\Omega})}{\mathrm{i}\widehat{\Omega}+f_\eta(\widehat{\Omega})}, \quad q_{d}=\sqrt{\frac{1-\eta}{1+\eta}}, \quad q_{c}=\frac{\eta\omega_0}{c} \\  
   & \begin{array}{c|c|c}
     & \textrm{discrete} & \textrm{continuous} \\
     \hline
     \widehat{\Omega} & \Omega\sqrt{1-\Omega^2/4} & \Omega \\ 
     f_\eta &  \eta+\sqrt{\eta^2-\widehat{\Omega}^2} &  (1+\eta)\left(\eta+\sqrt{\eta^2-\Omega^2}\right)
    \end{array}
    \end{split}
    \label{eq:R_theory}
\end{equation}
where $\Omega=\omega/\omega_0$,
and are depicted in \autoref{fig:R_theory} and \autoref{fig:decay_theory}.
It can be observed that in both the discrete and the continuous cases $R$ contains a square root expression, that once real, implies $|R|=1$, indicating the `no tunneling' state. This occurs at a turning point, which is denoted by the threshold frequency $\Omega_g$.
In the continuous system (solid curves) $\Omega_g=\eta$ is obtained, meaning $|R|<1$ for $\Omega<\Omega_g$. For $\Omega>\Omega_g$, $|R|$ is decreasing to an $\eta$-dependent nonzero value, enabling the tunneling.
In the discrete system (dashed lines), two threshold frequencies are obtained: a lower one $\Omega_{g-}=\sqrt{2}\sqrt{1-\mu}$ and an upper one $\Omega_{g+}=\sqrt{2}\sqrt{1+\mu}$, where $\mu=\sqrt{1-\eta^2}$ (due to the quartic relation resulting from $\widehat{\Omega}^2=\Omega^2(1-\Omega^2/4)$). For all $\eta>0$, $|R|=1$ for $\Omega<\Omega_{g-}$. 
For $\Omega>\Omega_{g-}$, $|R|$ begins to decrease below 1, indicating that tunneling becomes possible. Then $|R|$ sharply increases back to unity toward the upper limit $\Omega_{g+}$, and remains unity up to the discrete propagation limit $\Omega=2$.

The decay rate, which is frequency-independent in the discrete and continuous case, is depicted in \autoref{fig:decay_theory} as $q_d^n$ for the discrete chain (with $n$ being the unit cell number), and $e^{-q_cx}$ for the continuous system. Both $q_d$ and $q_c$ increase with $\eta$, which implies that for a higher $\eta$ the interface is darker. This indicates the trade-off with the reflection coefficient: a darker interface results in less energy transmitted through the non-Hermitian interface. 

To calculate the reflection coefficient and decay rates for the proposed active metamaterial with $N$ control sources, frequency domain simulations were performed using the FE model. The reflection coefficient was calculated via $R=\hat{p}_r/\hat{p}_i$, where $\hat{p}_r$ and $\hat{p}_i$ are the complex amplitudes of the reflected and incident plane waves at $x=0$.
\autoref{fig:R_fem_9} shows the simulated reflection coefficient for $N=9$ unit cells and $\eta$ ranging from $0.2$ to $0.8$. These results confirm the general trends predicted by the theory: at low frequencies, the metamaterial is strongly reflective due to the barrier induced by the non-Hermitian skin effect, manifesting at the interface at $x=0$. Above a certain frequency (which increases with increasing $\eta$), the reflection coefficient decreases, indicating the onset of the tunneling behavior.

A key difference between the FE simulation results and the theory is that the numerical results are oscillatory. 
The peak and dips that can be observed in the simulated reflection coefficient are resulting from constructive and destructive interference effects caused by scattering of sound waves at the finite metamaterial interface (the actively controlled part between the paddings). If a larger number of unit cells is considered, as shown in \autoref{fig:R_fem_21}, more peaks and dips are introduced.
Additionally, in both \autoref{fig:R_fem_9} and (d) a peak can be observed near the mechanical resonance frequency of the control sources $f_s = \SI{235}{\hertz}$. This indicates that the actuator self-dynamics cancellation via the $C_{Sd}$ term in \autoref{eq:controllers} is not fully accurate because of the extraction of the sound pressure $p_n$ at a single point at a distance of $e$ away from the diaphragm.

\begin{figure*}

\figline{
        \fig{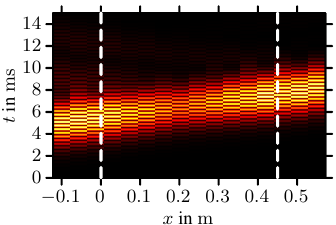}{57mm}{(a)}
        \label{fig:time_sim_eta_0_1_discrete}
        \fig{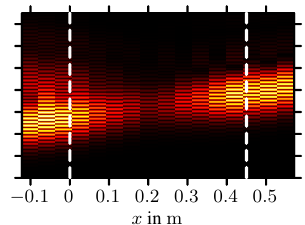}{52mm}{(b)}
        \label{fig:time_sim_eta_0_3_discrete}
        \fig{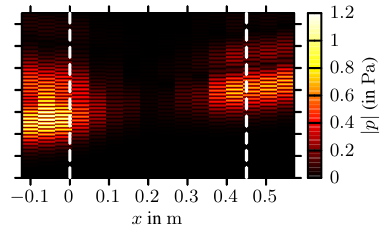}{66mm}{(c)}
        \label{fig:time_sim_eta_0_6_discrete}
    }
\figline{
        \fig{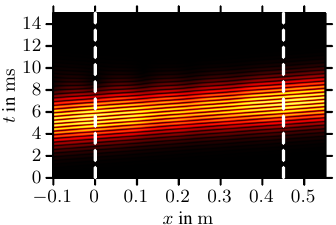}{57mm}{(d)}
        \label{fig:time_sim_eta_0_1_fem}
        \fig{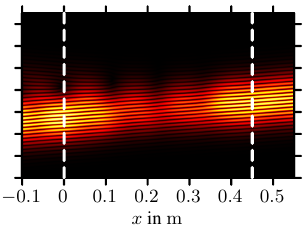}{52mm}{(e)}
        \label{fig:time_sim_eta_0_3_fem}
        \fig{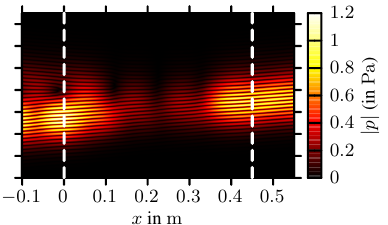}{66mm}{(f)}
        \label{fig:time_sim_eta_0_6_fem}
    }
\figline{
    \fig{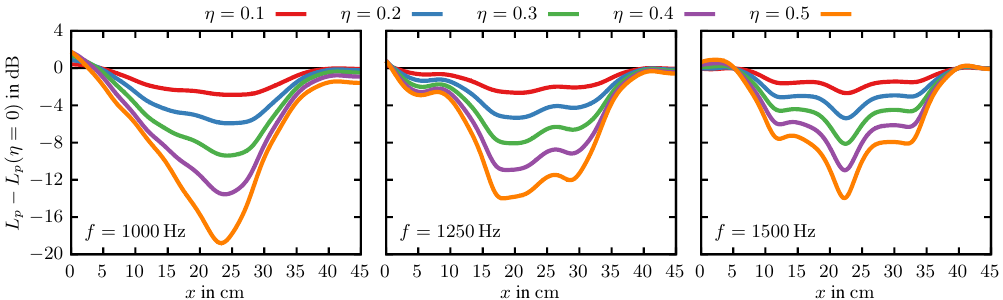}{6.69in}{(g)}
    \label{fig:time_sim_tunnel_fem}
}

\caption{(a)-(c) Time-dependent simulation results for the equivalent lattice model for $\eta=0.1$, 0.3, and 0.6, at $f=1500$ Hz. (d)-(f) The corresponding time-dependent FE simulation results for the controlled metamaterial. (g) Simulated closed-loop sound pressure levels $L_p$ (relative to the uncontrolled case with $\eta=0$) along the waveguide axis for $N=9$ active unit cells at different frequencies.}
\label{fig:time_sim}
\end{figure*}

\subsection{Tunneling of a wave packet}
\label{sec:numerical_tunneling}

Transient simulations were performed to investigate the propagation of a sine-modulated Gaussian pulse through the active metamaterial for different values of $\eta$. The FE simulations of the waveguide are compared to simulations of the discrete chain with equivalent acoustic parameters $M_0$ and $K_0$, as defined below \autoref{eq:constitutive}.
\autoref{fig:time_sim_eta_0_1_discrete}-(c) shows the simulation results for the discrete model with different $\eta$ values. The cell spacing $a$ was taken $5$ $\textrm{cm}$, as the spacing in the waveguide, whereas $S_w$ was taken as $H^2$ (as in the experimental setup described in \autoref{sec:experiment}).
\autoref{fig:time_sim_eta_0_1_fem}-(f) shows the corresponding FE simulation results of the waveguide model, indicating a good overall agreement between the models. The boundaries of the metamaterial are marked using the vertical dashed lines. In both cases, the tunneling can be seen to increase in strength with increasing $\eta$, as shown by the darker region in the middle of the metamaterial.

\autoref{fig:time_sim_tunnel_fem} shows additional frequency domain FE simulation results for the sound pressure level $L_p$ along the waveguide axis in closed-loop configuration. Three different frequencies above the tunneling threshold frequency of the metamaterial were tested.
The $L_p$ values are plotted relative to the sound pressure levels for the uncontrolled case ($\eta=0$) to illustrate the reduction of the sound pressure as the sound wave is tunneled through the metamaterial.
In general, a frequency dependence of the tunneling strength can be observed in these results.
At \SI{1000}{\hertz} in particular, $L_p$ decreases almost linearly along the metamaterial until it reaches a minimum close to the metamaterial center. $L_p$ then increases again and recovers to almost the same level as the incident wave. As expected from the theoretical considerations, the strength of the tunneling increases with increasing $\eta$, reaching a reduction by almost \SI{20}{\deci\bel} for $\eta=0.5$. It can also be observed that the sound pressure level values just in front of the metamaterial (at $x=0$) become elevated, which is due to the reflection at the interface, which becomes stronger for higher $\eta$ (see \autoref{fig:fem_reflection}).

\section{Experimental demonstration}
\label{sec:experiment}

To demonstrate a practical implementation of the proposed active metamaterial and verify the expected tunneling effect, acoustic measurements using a rectangular acoustic waveguide were conducted. \autoref{sec:experiment_setup} describes the general design and setup of the experiment. The measurement results are shown in \autoref{sec:experiment_results}, including a comparison to numerical results obtained from the FE model described in \autoref{sec:numerical}.

\subsection{Measurement setup}
\label{sec:experiment_setup}

\begin{figure*}
    \figline{
        \fig{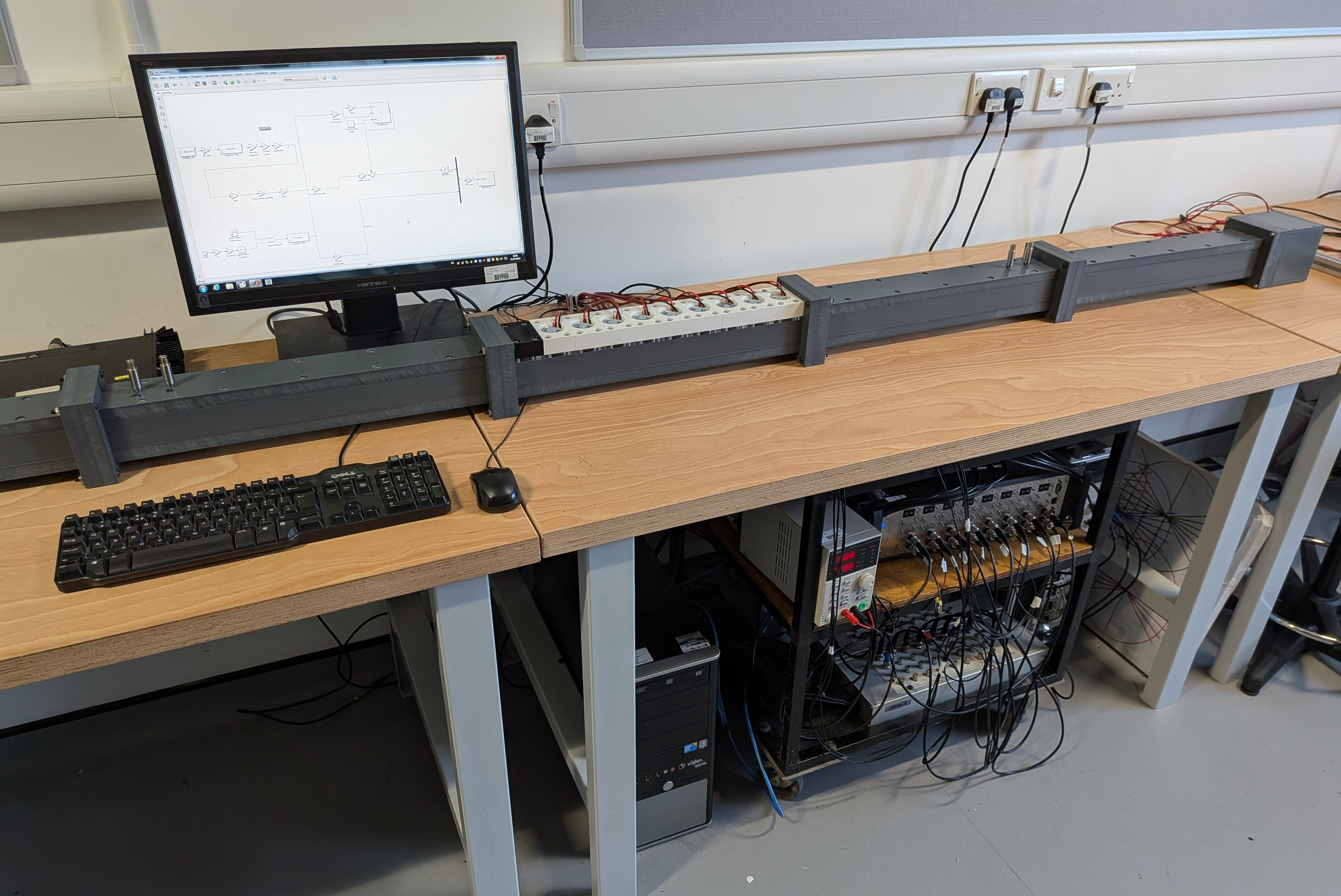}{2.1in}{(a)}
        \label{fig:exp_setup}
        \fig{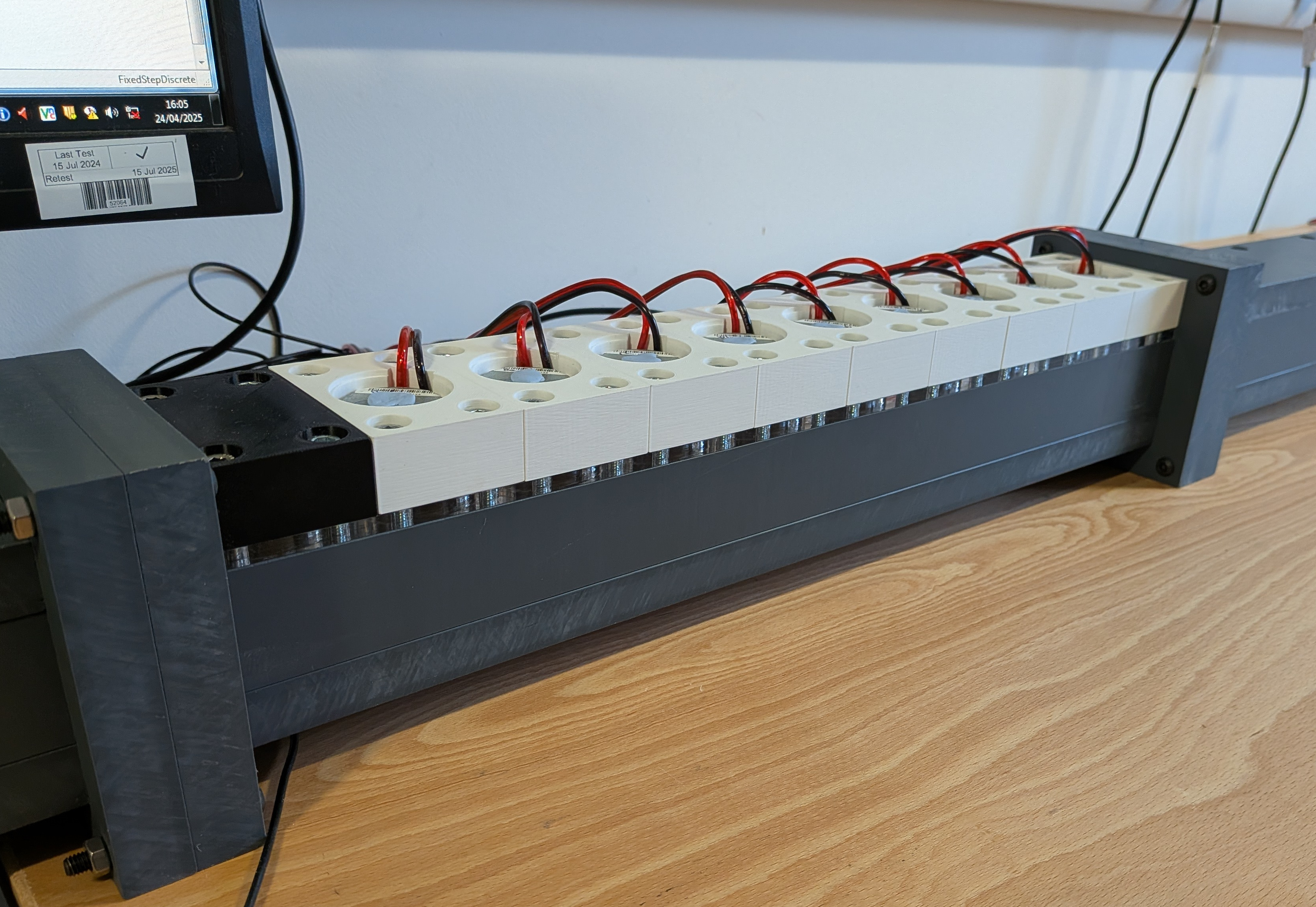}{2.04in}{(b)}
        \label{fig:exp_photo}
        \fig{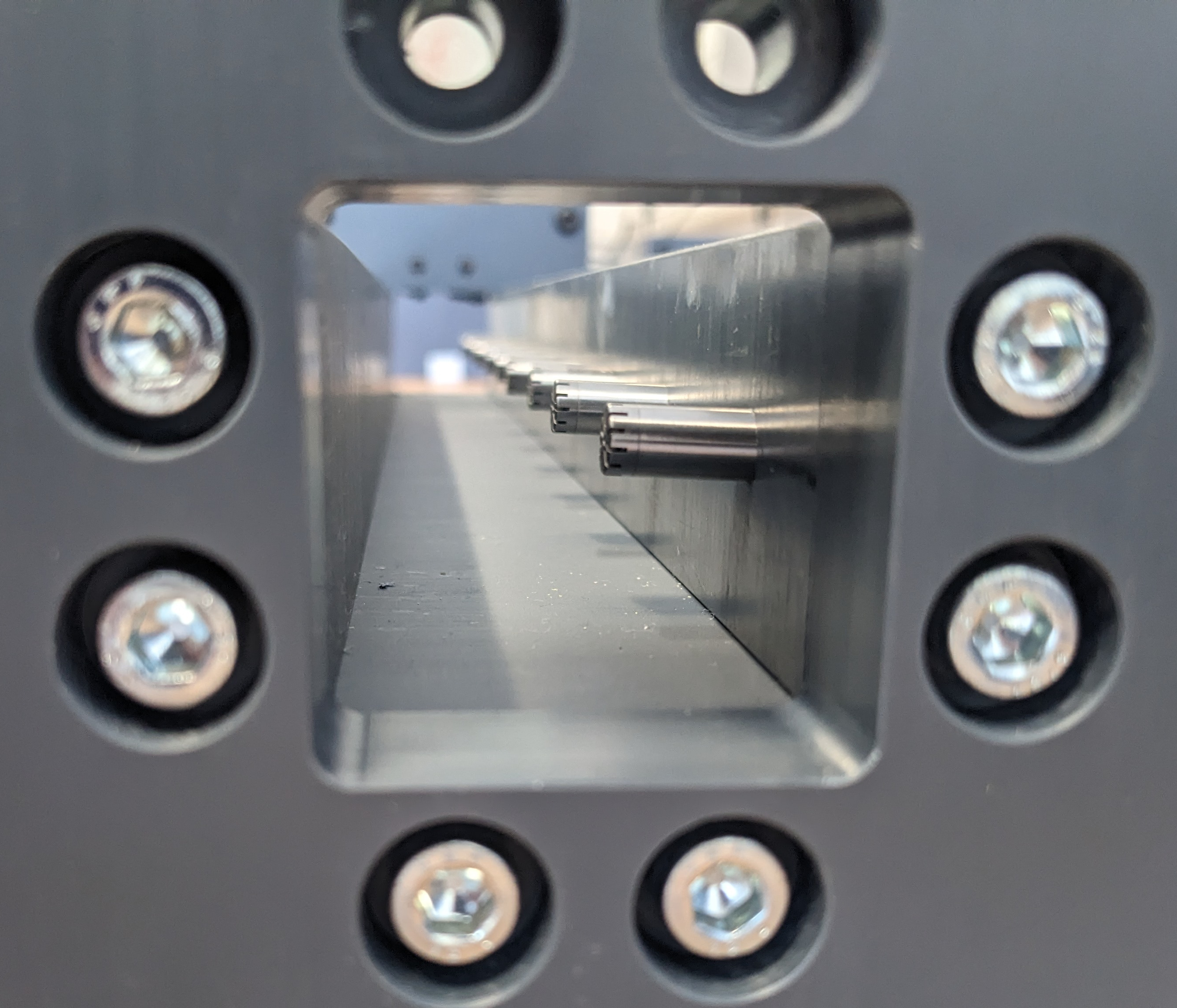}{1.638in}{(c)}
        \label{fig:exp_mics}
    }    
     \figline{
        \fig{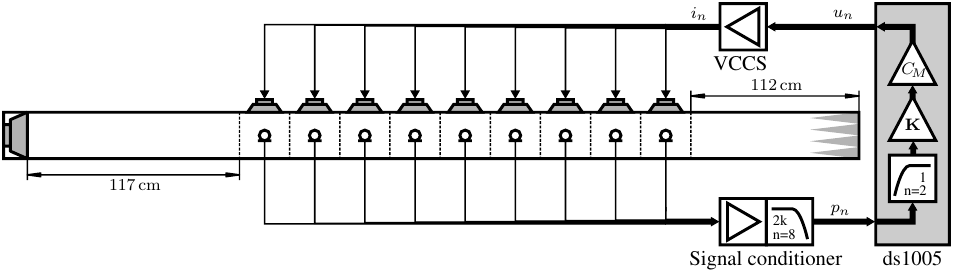}{5.5in}{(d)}
        \label{fig:exp_overview}
    }

    \caption{Experimental setup for the measurement of the tunneling effect in the proposed active metamaterial. (a) Overview of the whole waveguide used in the experiments. (b) Photograph of the active metamaterial with the back of the nine control sources visible. (c) Close-up view of the inside of the test section, showing the placement of the control microphones in the waveguide. (d) Sketch of the experimental implementation of the control setup.}
    \label{fig:experiment}
\end{figure*}

\autoref{fig:exp_setup} shows an overview of the whole waveguide used for testing of the active acoustic metamaterial in the center of the waveguide. The waveguide was made from \SI{20}{\milli\meter} thick polyvinylchlorid with a square \SI{40}{\milli\meter} $\times$ \SI{40}{\milli\meter} cross section.
At one end of the waveguide, a primary disturbance source was used to generate an acoustic wave signal propagating towards the metamaterial. The other end of the waveguide was almost anechoically terminated using absorptive material (polyester fibres).
Nine active metamaterial unit cells were 3D printed and attached to the wall of the waveguide, see \autoref{fig:exp_photo}. Each unit cell contained one loudspeaker (Peerless TC5FB00-04), with the nominal Thiele/Small parameters, provided by the manufacturer, given in \autoref{tbl:ts_parameters}, mounted flush to the waveguide wall. 1/4 inch free-field microphones (GRAS 40PL-10) were used as the control sensors in the unit cells. As shown in \autoref{fig:exp_mics}, each microphone was positioned on the axis of the waveguide in front of the center of each corresponding control source.

\autoref{fig:exp_overview} illustrates the control setup that was implemented in the experiment. The output signals from the microphones were passed through a signal conditioner, containing an amplifier and a low-pass filter (8th order Butterworth with a cut-off frequency of \SI{2}{\kilo\hertz}). A rapid digital control prototyping platform (dSPACE ds1005) was used to implement the control laws derived in \autoref{sec:metamaterial_design}, converting the pressure signals $p_n$ measured by the microphones into control signals $u_n$. The sampling frequency of the digital controller was set to \SI{16}{\kilo\hertz}. First, the pressure signals were passed through a 2nd order Butterworth high-pass filter with a cut-off frequency of \SI{1}{\hertz} to filter out any DC components in the microphone signals. The pressure signals were then multiplied with a matrix gain $\mathbf{K}$, turning the pressure signals into pressure difference signals as per \autoref{eq:Delta_pn}. Then, the control law according to \autoref{eq:u_rep_n} was applied, however with two key simplifications: the $C_{Sd}$ part was neglected and the $C_\eta$ part, which contains the full mechanical impedance of the driver (see \autoref{eq:controllers}), was simplified to only contain the moving mass term. This significantly simplifies the controller to a purely proportional controller with the gain $C_M = -\eta M_{ms} / (\rho_0 a \beta Bl)$. Note that this approximation is only valid for frequencies $\gg f_s = \SI{235}{\hertz}$ and reduces the stable range of $\eta$ values for the metamaterial.

The dSPACE system generates the control signals as voltages $u_n$ which correspond to the desired current signals $i_n$ driving the actuators and creating the tunneling effect. To convert the voltage outputs from the dSPACE into current signals, a custom-made voltage-controlled current source (VCCS) based on the improved Howland current pump circuit \citep{rivet2016broadband} with a gain of \SI{1}{\ampere\per\volt} was built, providing a one-to-one conversion of the voltage outputs into control current signals.

\subsection{Measurement results}
\label{sec:experiment_results}

Before performing the tunneling measurements, the sensitivities of all microphones were determined using a pistonphone microphone calibrator. These sensitivities were implemented in the dSPACE controller to convert the voltage input signals into acoustic pressure signals.

\begin{figure*}\figline{
        \fig{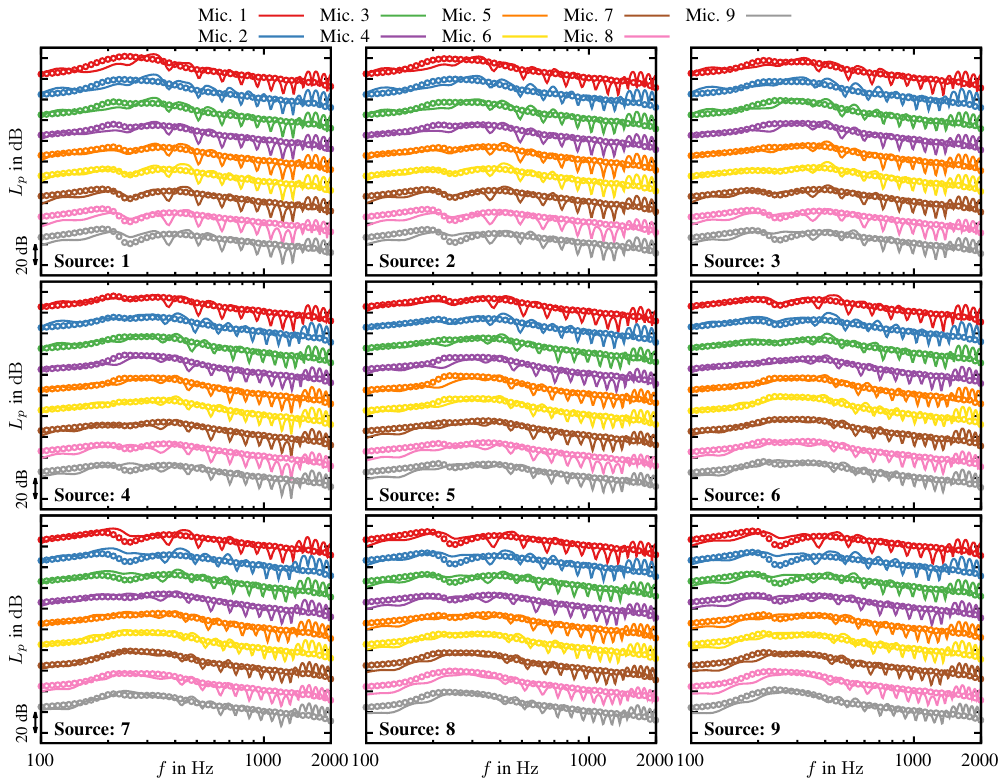}{6.2in}{}
    }
    \caption{Measured (lines) and simulated (circles) plant response magnitudes, shown as the sound pressure levels $L_p$ at each control microphone position for unit current signal excitation at each control source. For clarity, the datasets were vertically separated by \SI{20}{\deci\bel}.}
    \label{fig:plant_mag}
\end{figure*}

Plant response measurements were performed to validate the correct implementation of the experimental setup and the numerical model. These measurements were performed by playing band-limited white noise through each control source individually and recording the pressure signals at the control microphones.
\autoref{fig:plant_mag} shows the sound pressure levels at the nine microphone locations for sinusoidal current signals with amplitude $i_n = \SI{1}{\ampere}$ supplied to each control source individually. The curves represent the experimental results and the circles have been obtained from the FE model using a frequency domain study, with the 8th order Butterworth low-pass filter included in the numerical results. For clarity, the datasets have been separated vertically by \SI{20}{\decibel} in these plots.

Overall, the measured and simulated plant response magnitude agree reasonably well, with a general trend of maximum sound pressure level at the mechanical resonance frequency of the control source speakers ($f_s = \SI{235}{\hertz}$) and a roll-off towards higher frequencies due to the inertia of the moving mass of the speakers.
The most notable difference between the experiments and simulations is that the measured plant responses exhibit an oscillatory behavior which is not observed in the simulations. These deviations are primarily caused by the reflections of sound waves at the primary sound source, leading to (partially) standing waves in the experimental setup, which were not present in the simulation model due to the PML used at both ends of the waveguide. As shown in the supplementary material, including the measured waveguide termination impedances in the FE model can improve the agreement between the measurements and simulations significantly and therefore explains these deviations.

Another key difference between the measured and simulated plant response magnitudes can be observed at the higher frequency end, between $\approx$1000 and \SI{2000}{\hertz}, where the measured sound pressure levels are higher than the simulated ones. Comparing all curves, it can be noticed that the elevated sound pressure levels are consistent for each source and therefore linked to the characteristics of each individual control source. A likely explanation for this is that the control sources exhibit cone break-up within this frequency range, i.e.\ a deviation from the idealized piston-like behavior assumed in the FE model due to mechanical diaphragm resonances.

\begin{figure*}\figline{
        \fig{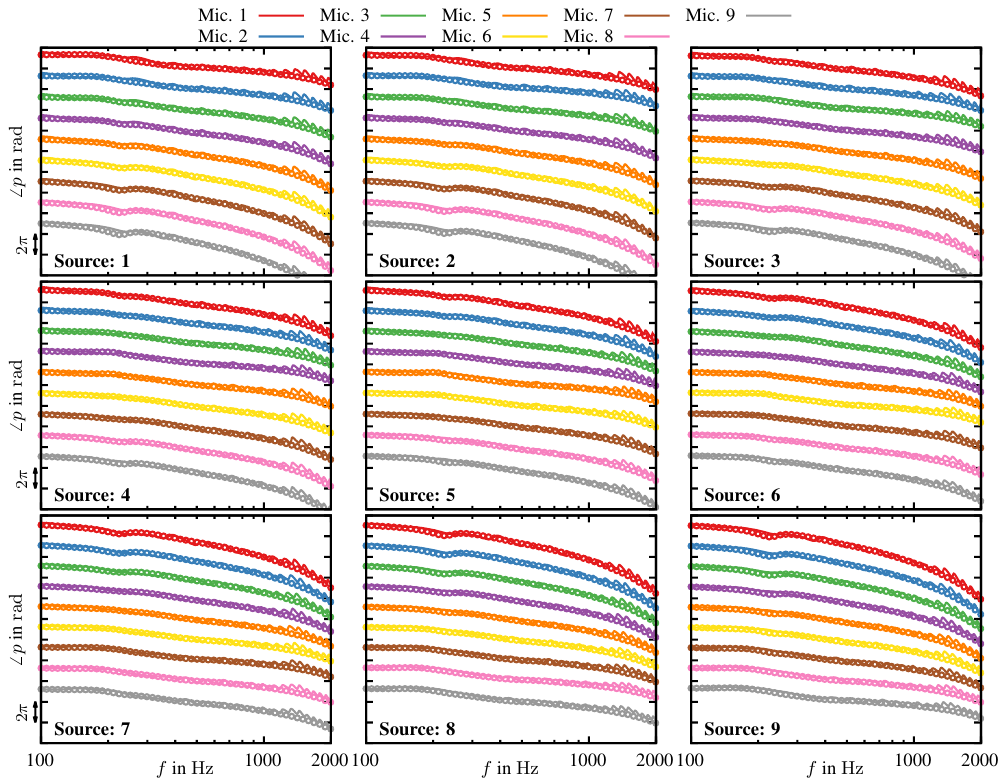}{6.2in}{}
    }
    \caption{Measured (lines) and simulated (circles) plant response phase angles at each control microphone position for unit current signal excitation at each control source. For clarity, the datasets were vertically separated by $2\pi$.}
    \label{fig:plant_arg}
\end{figure*}

The phase angles of the simulated and measured plant responses are shown in \autoref{fig:plant_arg}. For clarity, the microphone datasets have been separated vertically by $2\pi$.
Similarly to the plant response magnitude comparison, the comparison of the simulated and measured phase angles are generally in good agreement and the key deviations are caused by standing waves present in the measurements and the cone break-up of the control sources at frequencies above \SI{1000}{\hertz}.
Despite these deviations caused by the simplifications made in the numerical modeling, the good agreement between simulations and experimental results shows that the FE model captures the relevant physical mechanisms of the proposed active acoustic metamaterial with current-driven control sources.

\begin{figure*}
    \figline{
        \fig{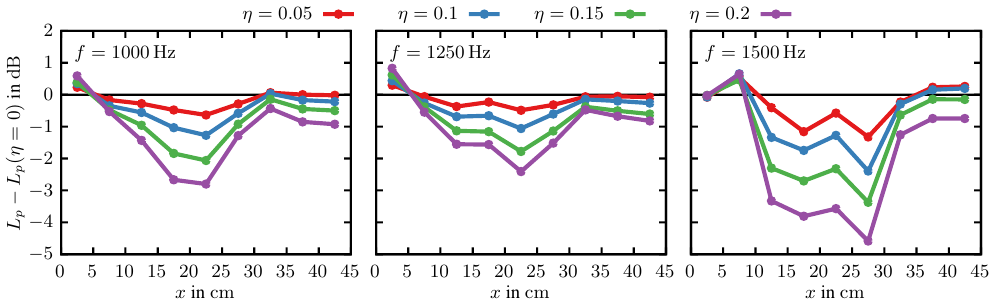}{6.2in}{(a)}
        \label{fig:exp_tnnl_N7}
        }
    \figline{
        \fig{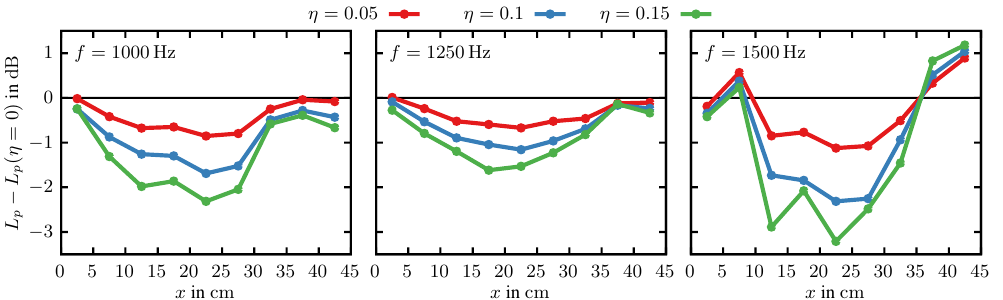}{6.2in}{(b)}
        \label{fig:exp_tnnl_N9}
    }
    \caption{Measured closed-loop sound pressure levels $L_p$ (relative to the uncontrolled case with $\eta=0$) at the control microphone locations for (a) $N=7$ and (b) $N=9$ active unit cells and sinusoidal primary source signals with different frequencies.}
    \label{fig:exp_tnnl}
\end{figure*}

For the experimental tunneling demonstration, sinusoidal signals with different frequencies were played through the primary disturbance source. For each frequency, the tunneling strength control parameter $\eta$ was increased in $0.05$ increments until the system became unstable. For each $\eta$ step, the resulting closed-loop pressure values $p_n$ at the control microphone locations were recorded.
\autoref{fig:exp_tnnl_N7} shows the closed-loop measurement results for $N=7$ active unit cells (corresponding to the unit cells $n=2,\ldots,8$) and three different frequencies: $f=\SI{1000}{\hertz}$, \SI{1250}{\hertz}, and \SI{1500}{\hertz}.
The results show the sound pressure levels $L_p$ at the control microphone locations relative to the measured sound pressure level values without control ($\eta=0$). This means, a value of \SI{0}{\deci\bel} corresponds to no change in sound pressure compared to the uncontrolled case and negative values indicate a reduction in sound pressure.
Four curves up to $\eta=0.2$ are shown, indicating that the system became unstable for $\eta=0.25$ and above. Although fairly low $\eta$ values are considered, the tunneling effect can be observed at all three frequencies shown in \autoref{fig:exp_tnnl_N7}. As expected from the theoretical and numerical models, the strength of the tunneling increases as $\eta$ is increased.
The tunneling strength also varies with frequency, with the strongest tunneling (up to \SI{5}{\deci\bel} sound pressure level reduction) achieved at \SI{1500}{\hertz}.

The tunneling measurement results for all $N=9$ active unit cells are shown in \autoref{fig:exp_tnnl_N9}.
In this case, the system remained stable only up to $\eta=0.15$, which indicates that the experimental implementation exhibits a trade-off between the overall number of active unit cells and the stability. There are various possible explanations for why the experimental system is more unstable than the numerical simulation model (for example: reflections at the waveguide terminations, variability of the control sources, delays introduced by the digital control system and anti-aliasing filters). Further investigations on the stability of the system and improvements of the experimental design are subject to future work.
Nevertheless, compared to the results for $N=7$ active unit cells, the tunneling region becomes wider, as the first and last unit cells now also contribute to the tunneling.
Additionally, a frequency dependence similar to that in the $N=7$ measurements can be seen, with the strongest tunneling observed at \SI{1500}{\hertz}.

In summary, the experimental results demonstrate the proposed tunneling effect under practical conditions, even with the simplified control law taking into account only the (nominal) moving mass of the control sources. Although the stability of the system limited the maximum $\eta$ values (and, consequently, the tunneling strength) to relatively low values, this could be improved by implementing the full control laws (as in \autoref{eq:u_rep_n}) or by applying sound absorbing material to the waveguide walls to minimize reflected sound waves.

\section{Conclusions}
\label{sec:conclusion}

In this contribution, a non-Hermitian active acoustic metamaterial design exhibiting a quantum-like tunneling behavior for acoustic waves was proposed. 
A discrete theoretical model of the metamaterial has been considered, consisting of a periodic chain of masses and non-reciprocal springs characterized by a non-reciprocity parameter $\eta$ (analogous to the electron creation and annihilation couplings in the original atomic chain). Based on this model, control laws were derived to enforce similar behavior in a hybrid discrete-continuous acoustic system.
The resulting proposed acoustic metamaterial consists of discrete volume velocity sources which are controlled, using pressure sensors, to give rise to the non-reciprocal behavior.

To demonstrate the functionality of the proposed metamaterial design, an FE model of a 2D acoustic waveguide with control sources located in the waveguide wall has been created. In the model, the control sources were modeled as current-controlled loudspeakers, via a lumped parameter model. Frequency domain simulations showed, as expected from the theory, that increasing the non-reciprocity parameter $\eta$ leads to a stronger tunneling through the metamaterial. Furthermore, it could be confirmed that the tunneling phenomenon occurs above a threshold frequency, which depends on $\eta$ and is given in \autoref{eq:R_theory}.
Time domain simulations were used to demonstrate that the proposed metamaterial is stable (under idealized conditions) for a large range of $\eta$ values and that transient signals, like a Gaussian wave packet, can be tunneled through the metamaterial with minor distortions of the waveform.

Finally, an experimental demonstration was realized using a square waveguide with nine control sources and a digital control system to implement a simplified control law for the non-Hermitian active metamaterial. Despite the non-idealized conditions (e.g., simplified control law, reflections at waveguide ends), the measurements confirmed the predicted tunneling behavior at different frequencies, with a tunneling strength of up to \SI{5}{\deci\bel}. 

The proposed active metamaterial design paves the way for various new ways of controlling sound waves in acoustic systems, without blocking the passage of sound or airflow like other active metamaterial designs. The numerical and experimental platforms developed in this contribution can also be used to investigate other types of non-Hermitian effects, such as supersonic sound wave propagation or group velocity acceleration.
Future work will investigate improvements of the experimental implementation of the active acoustic metamaterial to achieve stability for a wider range of $\eta$ values and stronger tunneling. An extension of the proposed design to higher dimensions (2D and 3D) will also be explored to study the tunneling effect in more complex acoustic wave fields.

\section*{Supplemental material}

See supplemental material at [URL will be inserted by AIP] for 
more details of the plant response measurements.

\begin{acknowledgments}
F.\ L. and J.\ T. have been partially supported by the UK's Engineering and Physical Sciences Research Council (EPSRC) through the 3rd funding call by the UK Acoustics Network Plus EP/V007866/1. L. S. and S. J. were supported by the Israel Science Foundation Grants 2177,2876/23.
The authors acknowledge the use of the IRIDIS High Performance Computing Facility, and associated support services at the University of Southampton, in the completion of this work.

\end{acknowledgments}

\section*{Author Declarations}
\subsection*{Conflict of Interest}
The authors have no conflicts to disclose.

\section*{Data Availability}

The data that support the findings of this study are
available from the corresponding author upon reasonable
request.

\bibliographystyle{jasanum2}

\end{document}


\title{\vspace{-2cm}Realizing non-Hermitian tunneling phenomena using non-reciprocal active acoustic metamaterials (supplemental material)}

\author[1]{Felix Langfeldt}
\author[1]{Joe Tan}
\author[2]{Sayan Jana}
\author[2]{Lea Sirota}
\affil[1]{Institute of Sound \& Vibration Research, University of Southampton, University Road, Highfield, Southampton, SO17 1BJ, United Kingdom}
\affil[2]{School of Mechanical Engineering, Tel Aviv University, Tel Aviv 69978, Israel}

\date{}

\maketitle

This document 
provides more detailed results for the plant response measurements of the experimental active acoustic metamaterial and a systematic comparison to the numerical simulation results, including an analysis of the effect of the tube terminations in the experimental setup.

\section{Detailed plant response measurement and simulation results}

\begin{figure}
    \centering
    \includegraphics{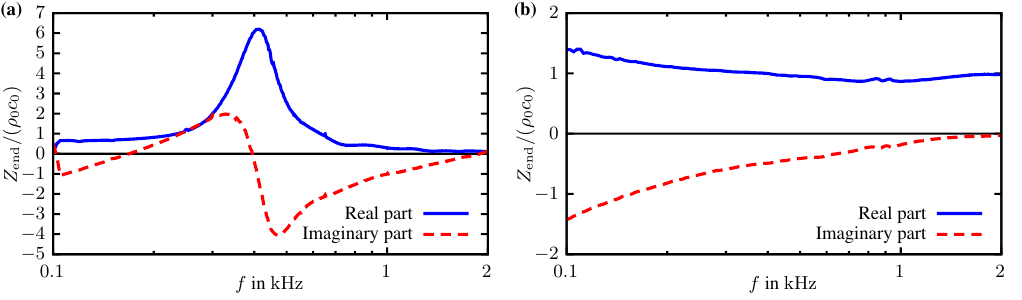}
    \caption{Measured specific acoustic impedances of the waveguide terminations. (a) Primary source end; (b) Anechoic termination end.}
    \label{fig:term_impedance}
\end{figure}

\begin{figure}
    \centering
    \includegraphics{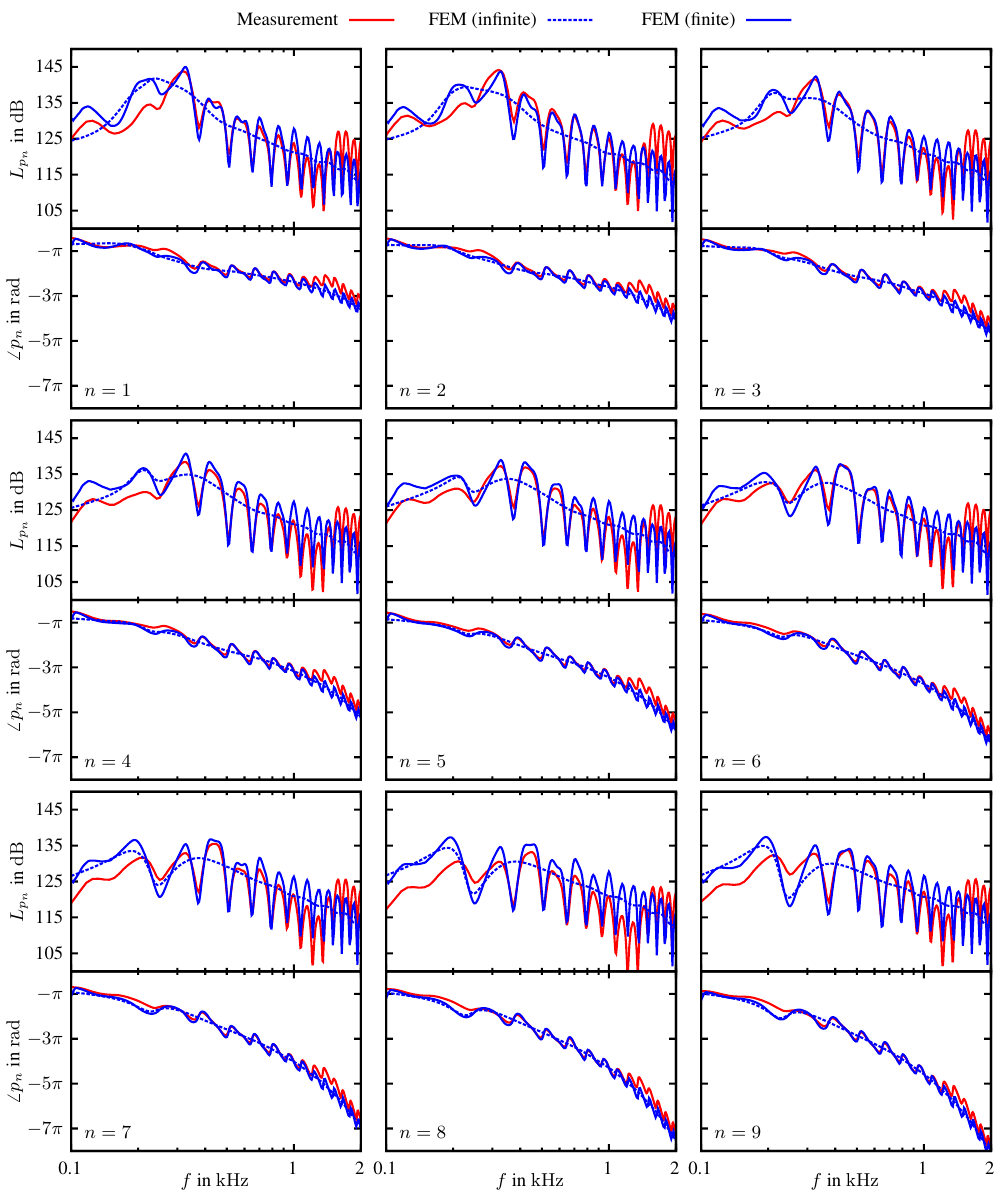}
    \caption{Measured and simulated plant responses of the active metamaterial for control source \#1.}
    \label{fig:plant_src1}
\end{figure}

\begin{figure}
    \centering
    \includegraphics{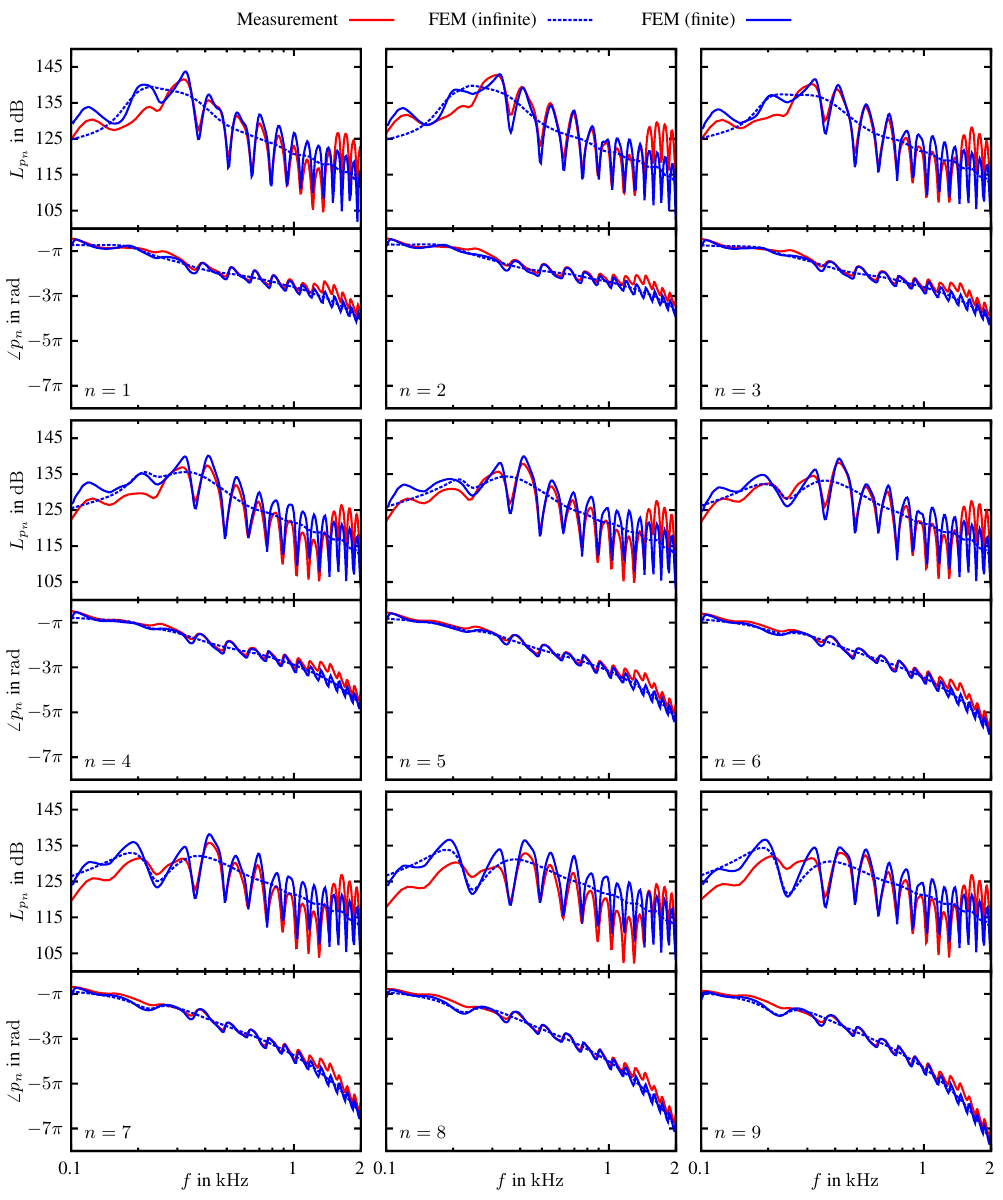}
    \caption{Measured and simulated plant responses of the active metamaterial for control source \#2.}
    \label{fig:plant_src2}
\end{figure}

\begin{figure}
    \centering
    \includegraphics{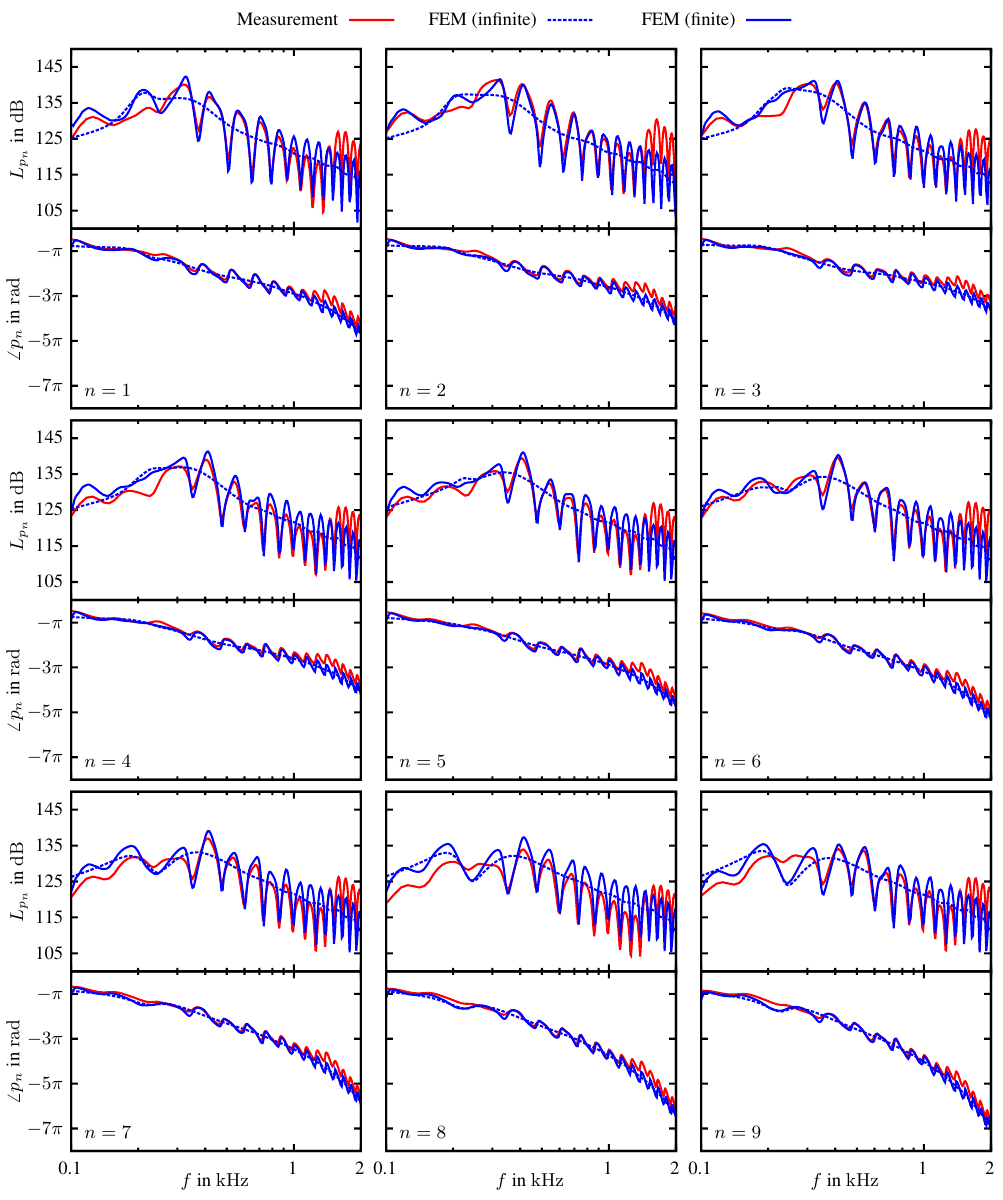}
    \caption{Measured and simulated plant responses of the active metamaterial for control source \#3.}
    \label{fig:plant_src3}
\end{figure}

\begin{figure}
    \centering
    \includegraphics{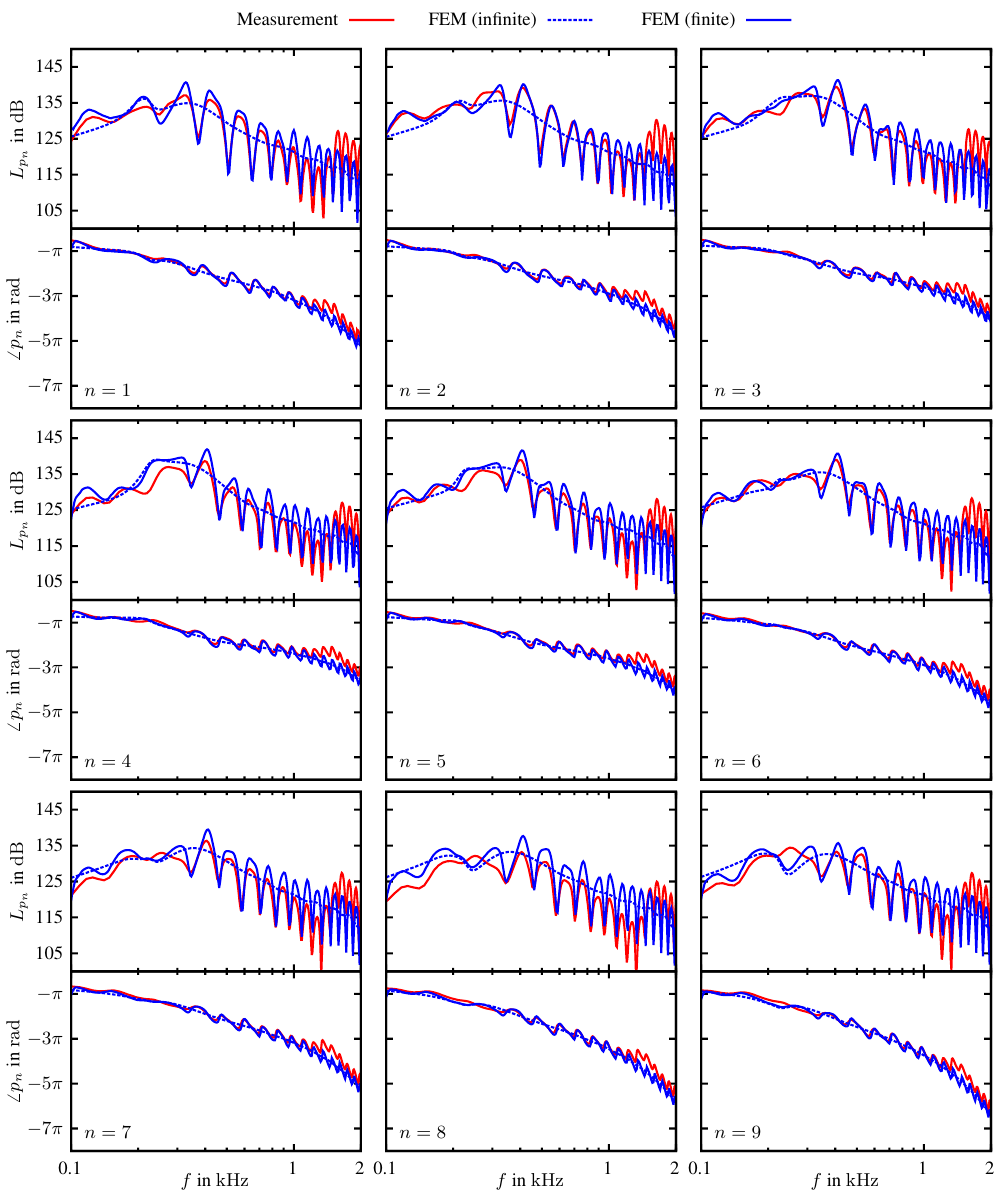}
    \caption{Measured and simulated plant responses of the active metamaterial for control source \#4.}
    \label{fig:plant_src4}
\end{figure}

\begin{figure}
    \centering
    \includegraphics{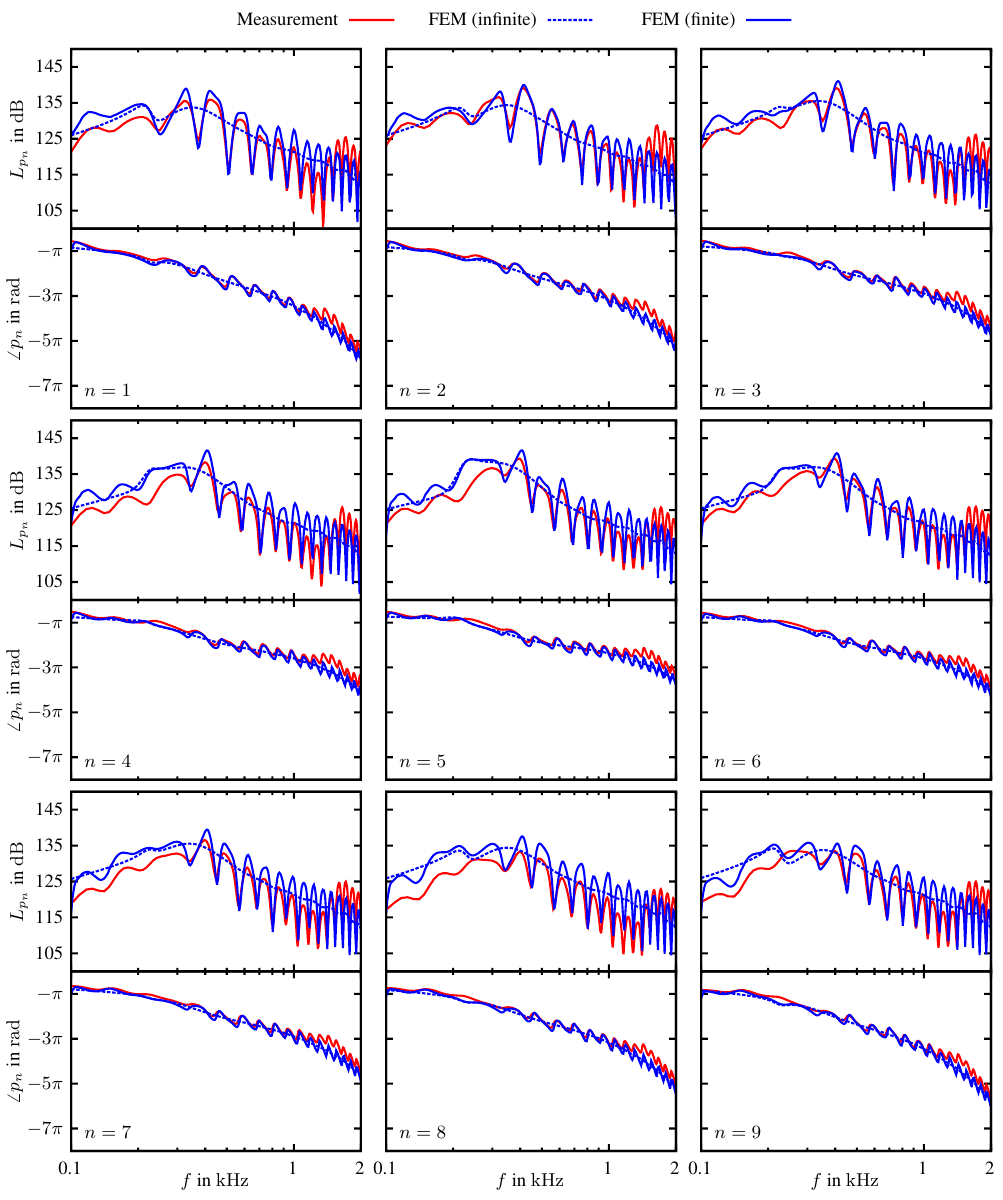}
    \caption{Measured and simulated plant responses of the active metamaterial for control source \#5.}
    \label{fig:plant_src5}
\end{figure}

\begin{figure}
    \centering
    \includegraphics{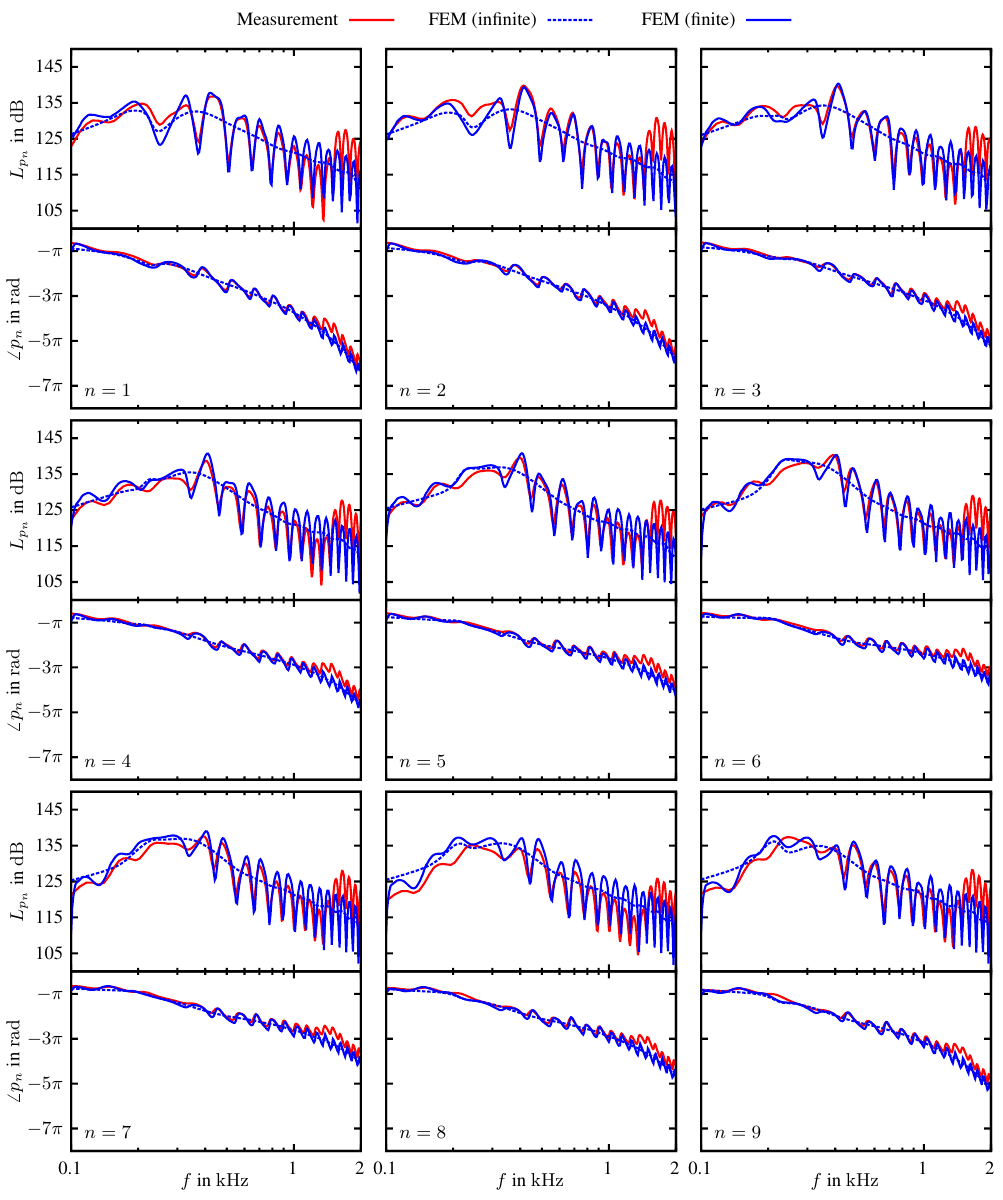}
    \caption{Measured and simulated plant responses of the active metamaterial for control source \#6.}
    \label{fig:plant_src6}
\end{figure}

\begin{figure}
    \centering
    \includegraphics{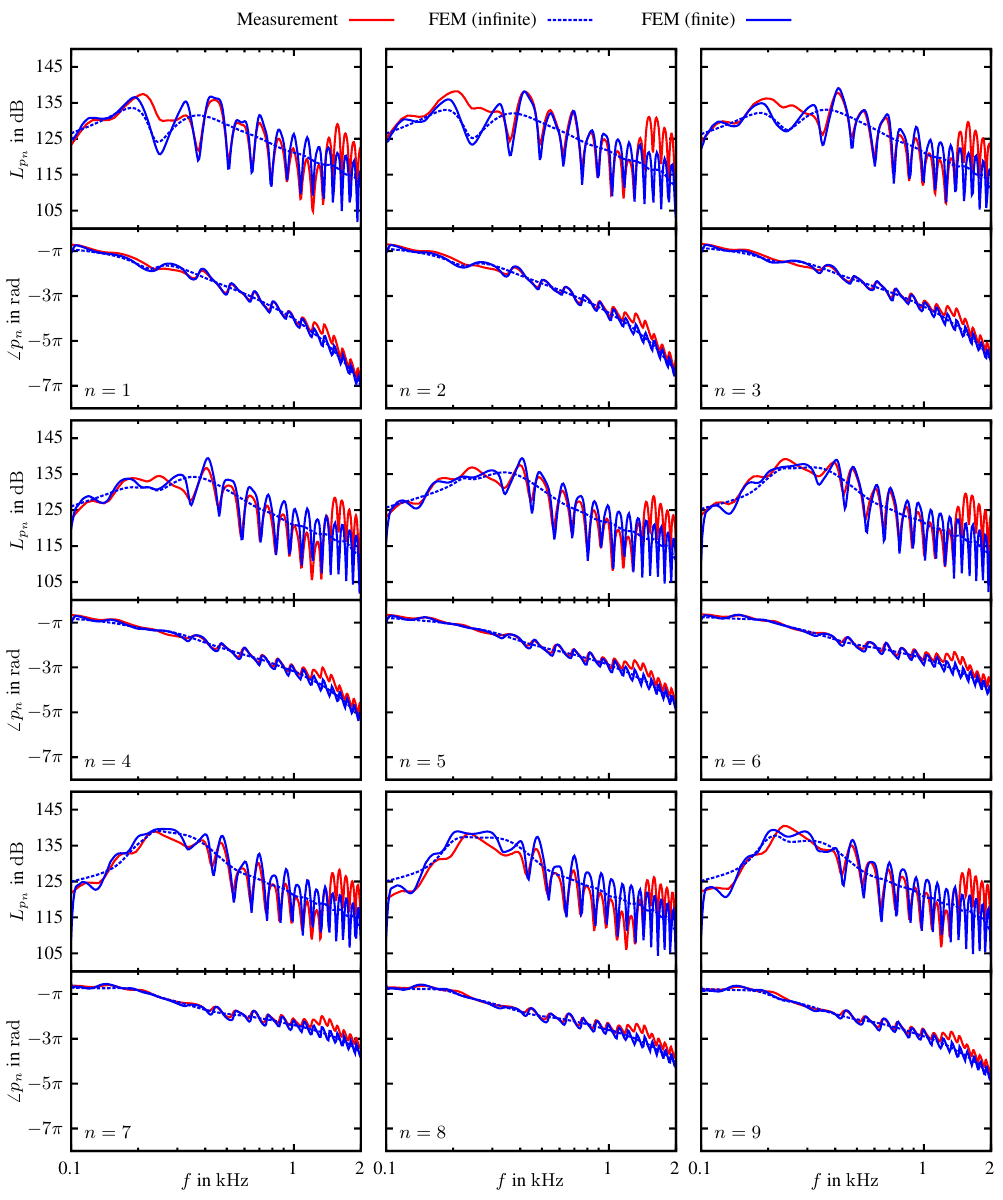}
    \caption{Measured and simulated plant responses of the active metamaterial for control source \#7.}
    \label{fig:plant_src7}
\end{figure}

\begin{figure}
    \centering
    \includegraphics{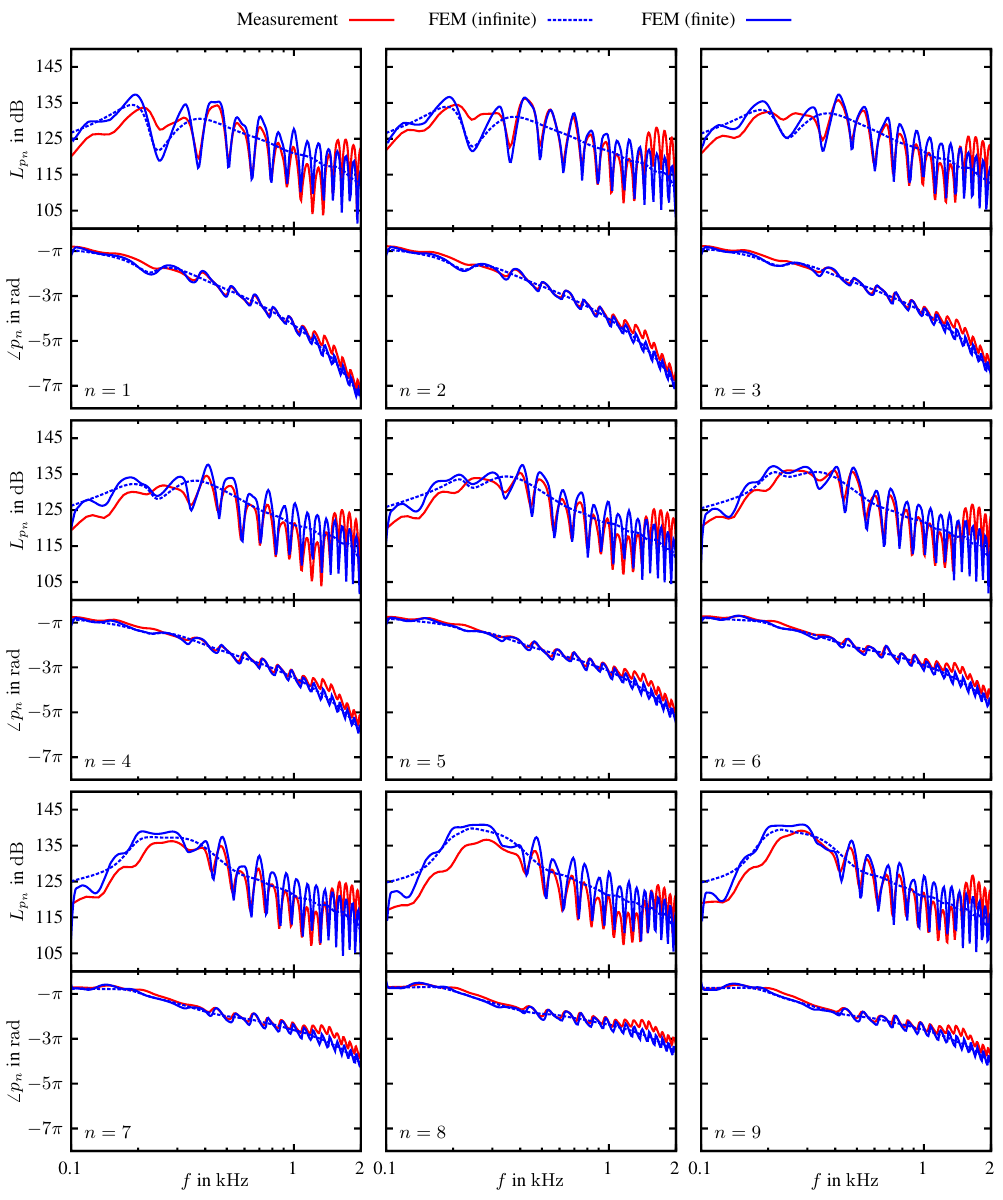}
    \caption{Measured and simulated plant responses of the active metamaterial for control source \#8.}
    \label{fig:plant_src8}
\end{figure}

\begin{figure}
    \centering
    \includegraphics{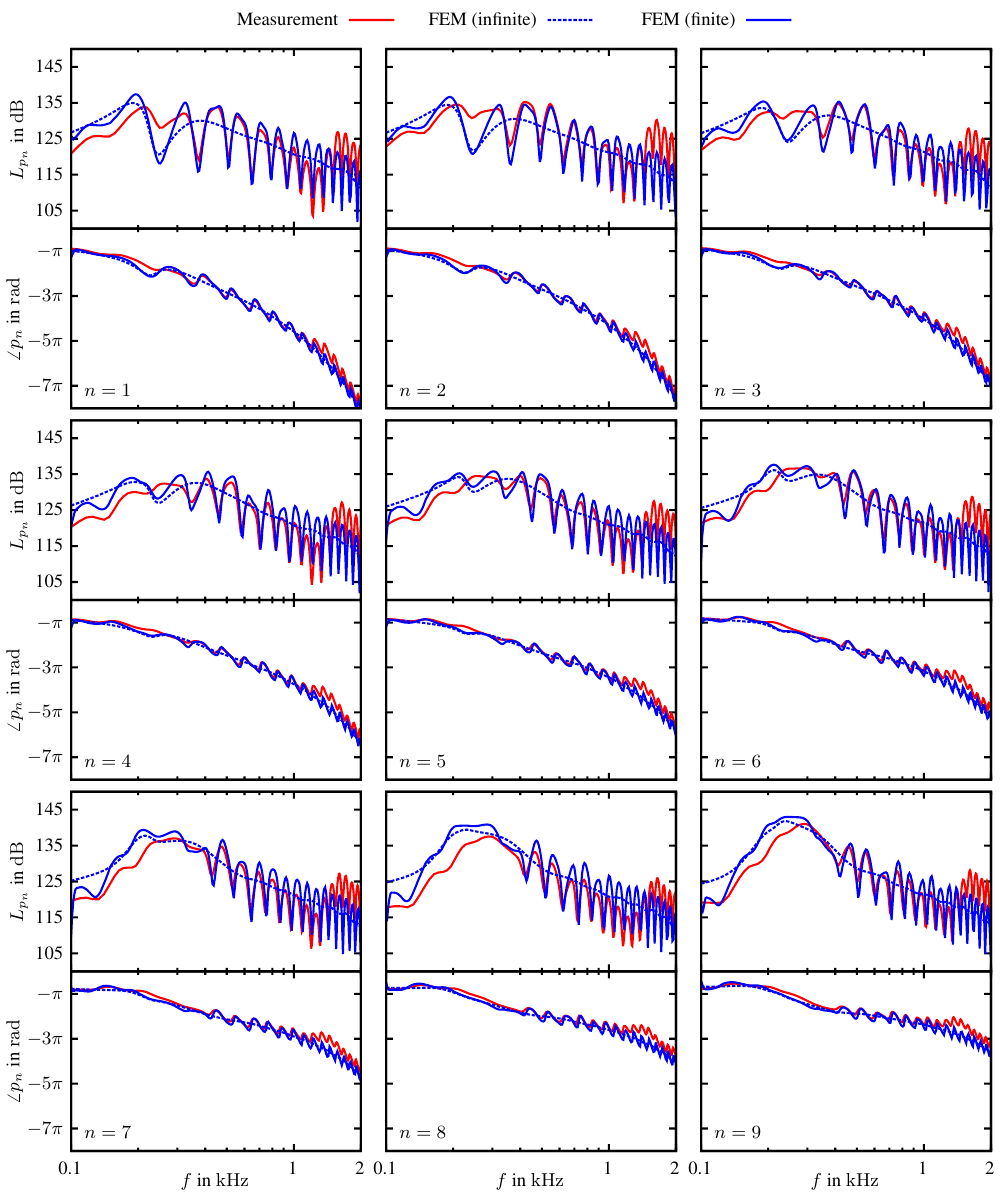}
    \caption{Measured and simulated plant responses of the active metamaterial for control source \#9.}
    \label{fig:plant_src9}
\end{figure}

\cref{fig:plant_src1,fig:plant_src2,fig:plant_src3,fig:plant_src4,fig:plant_src5,fig:plant_src6,fig:plant_src7,fig:plant_src8,fig:plant_src9} show the measured and simulated plant responses, corresponding to the pressure values $p_n$ ($n=1,\ldots,9$) resulting from a harmonic current signal with amplitude \SI{1}{\ampere} applied to the control sources 1 to 9, respectively. In each subplot, the top half shows the magnitude of the signal (in terms of the sound pressure level $L_{p_n}$), while the bottom half shows the phase angle of the pressure.
The red curves correspond to the measurement results and the blue curves have been obtained using the FEM model. The dotted curve represents the results obtained using the FEM model with non-reflecting terminations at both ends of the waveguide (as in the main paper), whereas the solid blue curve has been obtained by including the measured impedances of both ends of the waveguide (see \cref{fig:term_impedance}) in the model using impedance boundary conditions.

The comparison of the measured and simulated plant responses shows that the plant responses are much smoother when an infinite waveguide is considered. After incorporating the experimental waveguide termination impedances in the FEM model, the agreement between simulations and measurements becomes considerably better, which confirms that the oscillatory magnitude and phase curves in the measurements are primarily related to reflections of acoustic waves at the primary sound source. The other end of the waveguide was almost perfectly anechoic, as can be seen in \cref{fig:term_impedance}(b), where the measured impedance is very close to the characteristic impedance of the fluid, in particular at the higher end of the frequency range of interest.

